\documentclass[sn-apa]{sn-jnl}
\theoremstyle{plain}
\newtheorem{thm}{Theorem}[section]

\newtheorem{cor}[thm]{Corollary}

\theoremstyle{definition}

\newtheorem*{thm*}{Theorem}
\theoremstyle{remark}
\newtheorem*{rem}{Remark}

\theoremstyle{thmstyleone}%
\theoremstyle{thmstyletwo}%

\theoremstyle{thmstylethree}%

\usepackage{xcolor}
\usepackage{caption}
\usepackage{subcaption}

\begin{document}
\title[]{Multi-period static hedging of European options}

\author[1]{\fnm{Purba} \sur{Banerjee}}\email{purbab@iisc.ac.in}
\equalcont{These authors contributed equally to this work.}

\author[1]{\fnm{Srikanth} \sur{Iyer}}\email{skiyer@iisc.ac.in}
\equalcont{These authors contributed equally to this work.}

\author*[2]{\fnm{Shashi } \sur{Jain}}\email{shashijain@iisc.ac.in}
\equalcont{These authors contributed equally to this work.}

\affil[1]{\orgdiv{Department of Mathematics}, \orgname{Indian Institute of Science}, \orgaddress{\city{Bangalore}, \postcode{560012},  \country{India}}}

\affil*[2]{\orgdiv{Department of Management Studies}, \orgname{Indian Institute of Science}, \orgaddress{\city{Bangalore}, \postcode{560012},  \country{India}}}


\abstract{We consider the hedging of European options when the price of the underlying asset follows a single-factor Markovian framework. By working in such a setting, \cite{carr2014static} derived a spanning relation between a given option and a continuum of shorter-term options written on the same asset. In this paper, we have extended their approach to simultaneously include options over multiple short maturities. We then demonstrate a practical implementation of this extension with a finite set of shorter-term options to determine the hedging error using a Gaussian Quadrature method. A wide range of experiments are performed for both the \textit{Black-Scholes} and \textit{Merton Jump Diffusion} models, illustrating the comparative performance of the two methods.}   

%
%



\maketitle
\section{Introduction}
Financial crises over the past few decades have highlighted the growing importance of static and semi-static hedging strategies. More recently, the widespread COVID-19 pandemic emphasized the well-known phenomenon that every major financial crisis is always accompanied by numerous mini-crises. These crises cause asset prices to behave in an unpredictable fashion, triggering circuit breakers, trading halts, and increased risk aversion among investors. All of these factors make the application of dynamic hedging strategies complicated and often faulty. Consequently, static and semi-static hedging strategies offer an attractive alternative.

One of the pioneering works in this regard was by \cite{breeden1978prices}. They proved that for a given portfolio, the price of a $\$1$ claim received at a future date provided the portfolio's value is between two specified levels on that date, can be obtained explicitly from a second partial derivative of its call option pricing function. This was further elaborated by  \cite{green1987spanning} and  \cite{nachman1988spanning}, who show that a path-independent payoff can be hedged using a portfolio of standard options maturing with the claim. In spite of the strategy being robust to model misspecification, the class of claims that this static hedging strategy can hedge is fairly narrow.

In \cite{figlewski2018risk}, the author provides a detailed analysis of the literature surrounding risk-neutral densities, with a focus on U.S. equity options. A problem arises in the case of stocks with relatively limited option trading, usually listed over only a few strikes. The author explains the difficulties in such scenarios, for both parametric and non-parametric risk-neutral densities, where the distribution of available exercise prices may become fairly asymmetrical, thereby providing considerably sparse information about the density's tail on one side. To address such issues, for non-parametric cases, \cite{figlewski2008estimating} proposed fitting Generalized Extreme Value (GEV) distributions to the missing tails, by utilising the Fisher-Tippett Theorem, which proves that the remote (right) tail of any 
plausible choice for a returns density will converge to the form of a GEV tail. Further simplification of this approach on substitution of the Generalized Pareto distribution (GPD) can be found in \cite{birru2012anatomy}.

In their $1997$ paper, \cite{carr1997hedging} propose static replications of barrier options using vanilla options under the \cite{black1973valuation} environment. The necessity of continuous trading of the underlying is replaced by the necessity of trading options with a continuum of different strikes and is restricted to the Black Scholes ($BS$) model. 

In the recent past, \cite{carr2014static} extend the strategy to obtain an exact static hedging relation to hedge a long-term option with a continuum of short-term options, all sharing a common maturity. This theoretical result, when discretized using their approach, to include finitely many shorter-term options, results in strike points that are spread widely apart. The static hedging approach in  \cite{carr2014static} is restricted to a single maturity for the shorter-term options and they recommend the short-term maturity to be close to the target option's maturity.  For motivation as to why this static hedging approach is useful for risk management, we refer the reader to Section $1.2$ of \cite{carr2014static}. When the target maturity is long, short-term options with maturity closest to the target option are likely to be thinly traded over restricted strike ranges resulting in large hedging errors.

In this paper, we address the problem of static hedging of European options and present a valuation in the one-factor Markovian dynamics framework, where assimilation of shorter-term options helps reduce the hedging error in \cite{carr2014static}, by covering the unhedged risk arising out of a narrow range of strikes over which options of a single maturity are available. We extend the theoretical spanning relation obtained in \cite{carr2014static}. The hedging portfolio constitutes short-term options, all written over the same underlying asset as for the target option and with multiple choices for the shorter maturities. We obtain an exact theoretical spanning relation for the hedge portfolio in this case. This relation is then discretized using a  Gaussian Quadrature method to include short-term options with bounded strike ranges. Further, the portfolio is not just restricted to short-maturity call options but can include actively traded put options.

To summarise, the main contributions of our paper are as follows: 
\begin{enumerate}
    \item Extend the exact theoretical spanning relation in \cite{carr2014static} to include options not restricted to a common short maturity.
    \item Discretize the spanning relation using a Gaussian Quadrature ($GQ$) algorithm for practical application of our method to construct hedge portfolios with a finite number of options over multiple short maturities.
    \item Perform a comparative analysis of the performance of our method with the one in \cite{carr2014static} in each of the cases when the number of quadrature points, the short maturities, and the strike intervals are varied for the $BS$ model.
    \item Perform a comparative analysis of the performance of our method with the one in \cite{carr2014static}, in each of the cases when the number of quadrature points, the short maturities, the strike intervals, and the parameters governing the distribution of the stock price jumps are varied for the Merton Jump Diffusion ($MJD$) model.
    \item Study the performance of our method and the method in \cite{carr2014static}, in comparison to a Delta Hedging algorithm, throughout the duration of the hedge, using simulated stock paths, in both the $BS$ and $MJD$ models. 
\end{enumerate}

In related literature, \cite{bakshi1997empirical}, \cite{bakshi2003delta}, and \cite{dumas1998implied} use hedging performance to test different option pricing models. \cite{bakshi2000spanning} propose a general option-valuation strategy based on effective spanning using basic characteristic securities. \cite{renault1996option} consider optimal hedging under a stochastic volatility model. \cite{hutchinson1994nonparametric} propose to estimate the hedging ratio empirically using a nonparametric approach based on historical data. \cite{he2006calibration} and \cite{kennedy2009dynamic} set up a dynamic programming problem in minimizing the hedging errors under jump-diffusion frameworks and in the presence of transaction cost. Their method applied to only jump-diffusion frameworks and provided better performance than the standard dynamic hedging approach in the presence of transaction costs.  \cite{branger2006tractable} and \cite{branger2011tractable} propose robust dynamic hedges in pure diffusion models when the hedger knows only the range of the volatility levels but not the exact volatility dynamics.

For static payoff matching strategies, \cite{balder2006robust} consider discretization strategies for the theoretical spanning relation in \cite{carr2014static} when the strikes of the hedging options are pre-specified and the underlying price dynamics are unknown to the hedger. \cite{wu2016simple} propose an option hedging strategy that is based on the approximate matching of contract characteristics. The portfolio constructed using their approach required expanding along contract characteristics instead of focusing on risk. Hedging instruments close in characteristics to the target
contract must be chosen to minimize the expansion errors on characteristic differences. The portfolio includes a total of three short-maturity options over two short maturities and with the added assumptions that at all strikes and expiries, the calendar spreads and butterfly spreads are strictly positive, such that the \cite{dupire1994pricing} local volatility is well-defined and strictly positive. 

Among the most recent works, \cite{bossu2021functional} propose a functional analysis approach using
spectral decomposition techniques to show that exact payoff replication may be achieved with a discrete portfolio of special options. They discuss applications for fast pricing of vanilla options that may be suitable for large option books or high-frequency option trading, and for model pricing when the characteristic function of the underlying asset price is known. In their paper, \cite{lokeshwar2022explainable} develop neural networks for a regress-later-based Monte Carlo approach for pricing multi-asset discretely-monitored contingent claims. Their work demonstrates that any discretely monitored contingent claim- possibly high-dimensional and path-dependent— under Markovian and no-arbitrage assumptions, can be semi-statically hedged using a portfolio of short-maturity options.

The layout of the paper is as follows: Section \ref{sec2} provides a detailed explanation of the exact spanning relation as well as the discretization scheme given by \cite{carr2014static}. In Section \ref{sec3} we propose an exact multi-period static hedging relation to hedge a European call/put option using a continuum of options with finitely many different short maturities and discretize the approach by applying a method of Gaussian Quadrature to generate the optimal strikes and associated weights of the short-maturity options constituting the hedge portfolio. In Section \ref{sec5} we perform a series of numerical experiments for the $BS$ and $MJD$ models to provide a comparative analysis of the efficiency of our approach with \cite{carr2014static}. Section \ref{sec6} gives the conclusion and certain mathematical derivations for the theoretical results have been provided in Appendix.

\section{Hedging using options with a common short maturity}\label{sec2}
 We restrict our attention to a continuous-time one-factor Markovian setting and show how one can approximately hedge the risk of a European option by holding a finite number of shorter-term European options, all having a common maturity, as proved in \cite{carr2014static}. We begin by stating the assumptions and notations that we shall use throughout this paper, followed by some of the theoretical results that we shall use later to approximate the static hedge using a finite number of shorter-term options. The results that are presented here can be readily extended to the case of a European put option via put-call parity.
 \subsection{Assumptions and Notations}
 We assume the markets to be frictionless and have no-arbitrage. We use the standard notation of $S_t$ to denote the spot price of an underlying asset (for example, a stock or stock index), at time $t$. To be consistent with the assumptions as well as notations in \cite{carr2014static}, we further assume that the owners of this asset enjoy limited liability, which implies that $S_t\geq 0$ at all times and the continuously compounded risk-free rate is a constant, $r$ and a constant dividend yield, $\delta $. Our analysis is also restricted to the class of models for which the risk-neutral evolution of the stock price process is Markovian in terms of the filtration generated by the stock prices $S$ and the calendar time $t$.
 
 We shall use $C_t(K, T)$ to denote the time-$t$ value of a call option with strike price $K$ and expiry $T$. The probability density function of the asset price under the risk-neutral measure $\mathbb{Q}$, evaluated at the future price level $K$ and the future time $T$, conditional on the stock price starting at level $S$ at an earlier time $t$, is denoted by $q(S,t, K, T)$.
 
 One then obtains, as shown by \cite{breeden1978prices}, that the risk-neutral density is related to the second strike derivative of the call pricing function as follows
\begin{align}
    q(S,t,K,T)=e^{r(T-t)}\frac{\partial^2 C}{\partial K^2}(S,t,K,T).
\end{align}
This yields the fundamental result derived in  \cite{carr2014static}.

\begin{thm}\label{carrwuthm}
Under no-arbitrage and the Markovian assumption, the time-$t$ value of a European call option maturing at a fixed time $T\geq t$ relates to the time-$t$ value of a continuum of European call options of shorter maturity $u\in[t, T]$ by
\begin{align}
    \label{Carrwu}
C(S,t,K,T)=\int_{0}^{\infty}w(\mathcal{K})C(S,t,\mathcal{K},u)d\mathcal{K},
\end{align}
for all possible non-negative values of $S$ and at all times $t\leq u$. The weighting function $w(\mathcal{K})$ is given by
\begin{align}
\label{weights}
    w(\mathcal{K})=\frac{\partial^2 C}{\partial \mathcal{K}^2}(\mathcal{K},u,K,T).
\end{align}
\end{thm}
The static nature of the spanning relation (\ref{Carrwu}) is attributed to the fact that the option weights $w(\mathcal{K})$ are independent of $S$ and $t$. Hence, under the assumption of no-arbitrage, once the spanning portfolio is formed at the initial time $t$, no further re-balancing needs to be done until the maturity date of the options in the constructed hedge portfolio. The practical implication of Theorem \ref{carrwuthm} is that an investor can hedge the risk associated with taking a short position on a given option, by taking a static position in a continuum of shorter-term options.

It should also be observed that the weight $w(\mathcal{K})$ associated with the call option with maturity $u$ and strike $\mathcal{K}$, is proportional to the gamma that the target call option shall have at time $u$, provided the underlying asset price is $\mathcal{K}$ at that time point. Hence, as explained in \cite{carr2014static}, the bell-shaped curve, centered near the call option's strike price, that is projected by the gamma of a call option, implies that the highest weight is attributed to the options whose strikes are close to that of the target option. Moreover, as the common short maturity $u$ of the hedging portfolio approaches the target call option's maturity $T$, the underlying gamma becomes more concentrated around the strike price, $K$. So, taking the limit $u\rightarrow T$, the entire weight is found to be concentrated on the call option of strike $K$.

\subsection{Finite approximation using Gauss Hermite Quadrature}
The result in (\ref{Carrwu}) shows that a European call option can be hedged using a continuum of short-maturity calls. However, in practice,  investors cannot form a static portfolio involving a continuum of securities. Therefore in \cite{carr2014static}, the integral in (\ref{Carrwu}) is approximated using a finite sum, where the number of call options thereby used to construct the hedging portfolio is chosen in order to balance the cost from the hedging error with the cost from transacting in these options.

The integral in (\ref{Carrwu}) is approximated by a weighted sum of a finite number $(N)$ of call options at strikes $\mathcal{K}_j,j=1,2.., N,$ as follows
\begin{align}
\label{carrwu2}
  \int_{0}^{\infty}w(\mathcal{K})C(S,t,\mathcal{K},u)d\mathcal{K}\approx \sum_{j=1}^{N}\mathcal{W}_j C(S,t,\mathcal{K}_j,u),
\end{align}
where the strike points, $\mathcal{K}_j$, and their corresponding weights are chosen based on the Gauss-Hermite quadrature rule, as shown in \cite{carr2014static}.

As described in their paper, a map is constructed in order to relate the quadrature nodes and weights $\{x_j,w_j\}_{j=1}^N$ to the corresponding choice of option strikes, $\mathcal{K}_j$ and the portfolio weights, $\mathcal{W}_j$. The mapping function between the strikes and the quadrature nodes is given by
\begin{align}
\label{strike}
    \mathcal{K}(x)=Ke^{x\sigma\sqrt{2(T-u)}+(\delta-r-\sigma^2/2)(T-u)},
\end{align}
and the gamma weighting function under the Black-Scholes model is as follows
\begin{align*}
    \mathcal{W}(\mathcal{K})=\frac{\partial^2 C(\mathcal{K},u,K,T)}{\partial\mathcal{K}^2}=e^{-\delta(T-u)}\frac{n(d_1)}{\mathcal{K}\sigma\sqrt{T-u}},
\end{align*}
where $n(.)$ denotes the pdf of a standard normal random variable and $d_1$ is given by
\begin{align*}
    d_1 = \frac{ln(\mathcal{K}/K) + (r - \delta + \sigma^2/2)(T - u)}{\sigma\sqrt{T-u}}.
\end{align*}
 
Finally, using the Gauss-Hermite quadrature $\{w_j,x_j\}_{j=1}^N$ and the map (\ref{strike}), one obtains the respective strike points, $\mathcal{K}_j, j=1,2,..N,$ 
and  the associated portfolio weights are given by
\begin{align}
\label{CWweights}
    \mathcal{W}_j=\frac{\mathcal{W}(\mathcal{K}_j)\mathcal{K}'_j(x_j)}{e^{-x^2_j}}w_j=\frac{\mathcal{W}(\mathcal{K}_j)\mathcal{K}_j\sigma\sqrt{2(T-u)}}{e^{-x^2_j}}w_j.
\end{align}

\section{Multi-period static hedging approach}
\label{sec3}
In this section, we modify equation (\ref{Carrwu}) to obtain an exact spanning relation using options with multiple short maturities, over bounded strike ranges.  The corresponding finite-sum approximations of the hedging integrals are then obtained by the application of Gaussian and Gauss-Laguerre Quadrature rules. The point of contrast between the Gauss Hermite and the Gaussian Quadrature rule lies in the fact that while the former is a finite approximation method for an integral on an infinite domain, the latter serves as an approximation for a definite integral on a bounded interval. 

Our first job now is to define the Gaussian Quadrature rule for our hedging problem and then apply it accordingly for our numerical experiments. A detailed explanation of the Gaussian Quadrature rule has been provided in the Appendix and \cite{davis2007methods}.
\subsection{Hedging using options with multiple short maturities}\label{multi}
In practice, there are finitely many actively traded options with maturity $u_1$, which have strikes in the range $[K_{11},K_{12}]$ and equation (\ref{Carrwu}) is essentially approximated as follows
 \begin{align}
\label{gaussquad}
C(S,t,K,T)\approx\int_{K_{11}}^{K_{12}}w(\mathcal{K})C(S,t,\mathcal{K},u)d\mathcal{K}\approx \sum_{j=1}^{N}\mathcal{W}_j(\mathcal{K}_j) C(S,t,\mathcal{K}_j,u),
\end{align}
where $\mathcal{K}_j$'s are the strikes corresponding to the options with maturity $u_1$ and $\mathcal{W}_j(\mathcal{K}_j)$'s are the corresponding weights of the short-term options that one needs to hold in their portfolio. These are obtained by a direct application of the Gaussian Quadrature rule to the integral given in equation (\ref{gaussquad}).

In practice, at any given time $t$, prior to the maturity $T$ of the target option, options over multiple shorter maturities are available. Further, the approximation in (\ref{gaussquad}) excludes a wide range of strike points, $[0, K_{11}]\cup [K_{12},\infty]$.This entails an error when compared to the original formula (\ref{Carrwu}).
 
 However, there would be frequently traded options of other multiple short maturities that may be available at time $t$. We show how these options can be included in the hedge portfolio to partially compensate for the error incurred by only using options over a restricted strike range $[K_{11}, K_{12}]$.

We illustrate the procedure for including options of maturity $u_2$ and formulate a hedging scheme that gives a better approximation than the one involving a single maturity $u_1$. We begin by rewriting the equation 
(\ref{Carrwu}) as follows
 \begin{align}
 \begin{split}
     \label{hedgeint1}C(S,t,K,T) &=\int_{K_{11}}^{K_{12}}w(\mathcal{K}_1)C(S,t,\mathcal{K}_1,u_1)d\mathcal{K}_1 +\int_{0}^{K_{11}}w(\mathcal{K}_1)C(S,t,\mathcal{K}_1,u_1)d\mathcal{K}_1\\
&+\int_{K_{12}}^{\infty}w(\mathcal{K}_1)C(S,t,\mathcal{K}_1,u_1)d\mathcal{K}_1.
\end{split}
 \end{align}
 
Using (\ref{Carrwu}), with $T$ being replaced by $u_1$ and $u_1$ being replaced by $u_2$, we can write $C(S,t,\mathcal{K}_1,u_1)$ as
 \begin{align}
 \begin{split}
 \label{int1}
 C(S,t,K,T) &=\int_{K_{11}}^{K_{12}}w(\mathcal{K}_1)C(S,t,\mathcal{K}_1,u_1)d\mathcal{K}_1\\
&+\int_{0}^{K_{11}}w(\mathcal{K}_1)\left(\int_{0}^{\infty}w_2(\mathcal{K}_2,\mathcal{K}_1)C(S,t,\mathcal{K}_2,u_2)d\mathcal{K}_2\right)d\mathcal{K}_1 \\
&+\int_{K_{12}}^{\infty}w(\mathcal{K}_1)\left(\int_{0}^{\infty}w_2(\mathcal{K}_2,\mathcal{K}_1)C(S,t,\mathcal{K}_2,u_2)d\mathcal{K}_2\right)d\mathcal{K}_1,
\end{split}
 \end{align}
 where
 \begin{align*}
     w_2(\mathcal{K}_2,\mathcal{K}_1)=\frac{\partial^2 C}{\partial {\mathcal{K}^2_2}}(\mathcal{K}_2,u_2,\mathcal{K}_1,u_1)~~\text{and, $0\leq t \leq u_2<u_1<T$}.
 \end{align*}

 Now, changing the order of integration in (\ref{int1}) yields
 \begin{align*}
&\int_{0}^{K_{11}}w(\mathcal{K}_1)C(S,t,\mathcal{K}_1,u_1)d\mathcal{K}_1+\int_{K_{12}}^{\infty}w(\mathcal{K}_1)C(S,t,\mathcal{K}_1,u_1)d\mathcal{K}_1\\ &=\int_{0}^{\infty}\left(\int_{0}^{K_{11}}w(\mathcal{K}_1)w_2(\mathcal{K}_2,\mathcal{K}_1)d\mathcal{K}_1\right)C(S,t,\mathcal{K}_2,u_2)d\mathcal{K}_2\\
&+\int_{0}^{\infty}\left(\int_{K_{12}}^{\infty}w(\mathcal{K}_1)w_2(\mathcal{K}_2,\mathcal{K}_1)d\mathcal{K}_1\right)C(S,t,\mathcal{K}_2,u_2)d\mathcal{K}_2.
 \end{align*}

We are now ready to state the main result of this paper.
\begin{thm}\label{theorem}
     Under no-arbitrage and the Markovian assumption, the time- $t$ value of a European call option maturing at a fixed time $T>t$ relates to the time-$t$ value of a continuum of European call options having shorter maturities  $0\leq t\leq u_2<u_1\leq T$ by
\begin{align}
    \label{hedgeint4}
\begin{split}
C(S,t,K,T)&=\int_{K_{11}}^{K_{12}}w(\mathcal{K}_1)C(S,t,\mathcal{K}_1,u_1)d\mathcal{K}_1 +\int_{0}^{\infty}\tilde{w}_2(\mathcal{K}_2)C(S,t,\mathcal{K}_2,u_2)d\mathcal{K}_2,
\end{split}
\end{align}
with weights
\begin{align}
\label{GQweights}
w(\mathcal{K}_1)=\frac{\partial^2 C}{\partial \mathcal{K}_1^2}(\mathcal{K}_1,u_1,K,T),
\end{align}

 \begin{align}
     \label{modifiedwght1}
\tilde{w}_2(\mathcal{K}_2)=\int_{0}^{K_{11}}w(\mathcal{K}_1)w_2(\mathcal{K}_2,\mathcal{K}_1)d\mathcal{K}_1+
\int_{K_{12}}^{\infty}w(\mathcal{K}_1)w_2(\mathcal{K}_2,\mathcal{K}_1)d\mathcal{K}_1,
 \end{align}
 where
\begin{align*}
     w_2(\mathcal{K}_2,\mathcal{K}_1)=\frac{\partial^2 C}{\partial {\mathcal{K}^2_2}}(\mathcal{K}_2,u_2,\mathcal{K}_1,u_1),
 \end{align*}
 and $[K_{11},K_{12}]$ denotes the range of  strikes available at initial time corresponding to the options with maturity $u_1$.
\end{thm}
\begin{rem}
    \begin{enumerate}
         \item Equation 5 is an exact spanning relation involving options with maturities $u_1$ and $u_2$, respectively.
        \item  Equation (\ref{hedgeint4}) allows the investor to incorporate the options with short maturity $u_1$, available in the range $[K_{11}, K_{12}]$ while imposing no restrictions on the strike range corresponding to the options with shorter maturity $u_2$.
    \end{enumerate}
\end{rem}
 
Iterating the above procedure yields the following Corollary that allows us to construct static hedge portfolios with short maturity options maturity times take values in a finite set.
 \begin{cor}\label{corollary}
      Under no-arbitrage and the Markovian assumption, the time- $t$ value of a European call option maturing at a fixed time $T>t$ relates to the time-$t$ value of a continuum of European call options having shorter maturities $0\leq t \leq u_n<...<u_2<u_1\leq T$ by
\begin{align*}
C(S,t,K,T)&=\int_{K_{11}}^{K_{12}}w(\mathcal{K}_1)C(S,t,\mathcal{K}_1,u_1)d\mathcal{K}_1 +\int_{K_{21}}^{K_{22}}\tilde{w}_2(\mathcal{K}_2)C(S,t,\mathcal{K}_2,u_2)d\mathcal{K}_2\\
&+...+\int_{0}^{\infty}\tilde{w}_n(\mathcal{K}_n)C(S,t,\mathcal{K}_n,u_n)d\mathcal{K}_n,
\end{align*}
with
\begin{align*}
    \tilde{w}_{i}(\mathcal{K}_i)&=\int_{0}^{K_{i-1,1}}\tilde{w}_{i-1}(\mathcal{K}_{i-1})w_i(\mathcal{K}_i,\mathcal{K}_{i-1})d\mathcal{K}_{i-1}\\
    &+\int_{K_{i-1,2}}^{\infty}\tilde{w}_{i-1}(\mathcal{K}_{i-1})w_i(\mathcal{K}_i,\mathcal{K}_{i-1})d\mathcal{K}_{i-1},~~i=2,....,n,
\end{align*}
and
\begin{align*}
     w_i(\mathcal{K}_i,\mathcal{K}_{i-1})=\frac{\partial^2 C}{\partial {\mathcal{K}^2_i}}(\mathcal{K}_i,u_i,\mathcal{K}_{i-1},u_{i-1}),
 \end{align*} 
 where $[K_{i,1},K_{i,2}]$ denotes the range of  strikes available at initial time corresponding to the options with maturity $u_i, i=1,2..,n$.
 \end{cor}
 \begin{rem}
 \begin{enumerate}
     \item In a real-world scenario, actively traded options with maturity $u_n$ would be available for strikes over a bounded interval $[K_{n,1},K_{n,2}]$. Taking this into account, one obtains the final expression of the hedging portfolio as
 \begin{align}
 \begin{split}
C(S,t,K,T)&=\int_{K_{11}}^{K_{12}}w(\mathcal{K}_1)C(S,t,\mathcal{K}_1,u_1)d\mathcal{K}_1
+\int_{K_{21}}^{K_{22}}\tilde{w}_2(\mathcal{K}_2)C(S,t,\mathcal{K}_2,u_2)d\mathcal{K}_2\\
&+.....+\int_{K_{n,1}}^{K_{n,2}}\tilde{w}_n(\mathcal{K}_n)C(S,t,\mathcal{K}_n,u_n)d\mathcal{K}_n+\epsilon,
\end{split}
 \end{align}
 where
 \begin{align*}
     \epsilon =\int_{[0,K_{n,1}]\cup [K_{n,2}, \infty]}\tilde{w}_n(\mathcal{K}_n)C(S,t,\mathcal{K}_n,u_n)d\mathcal{K}_n, 
 \end{align*}
 denotes the approximation error.
 \item Restricting to the case of two short maturities, $u_1$ and $u_2$, as done for our numerical experiments in Section \ref{sec5}, we would like to highlight the following important observations. These can be readily generalized for the case of finitely many short maturities $0<u_n<...<u_2<u_1$.
 \begin{enumerate}
     \item Our hedge is valid till the short maturity $u_2$, after which the agent can decide to continue with their initial position in the options with maturity $u_1$, given by weights (\ref{GQweights}), over the remaining duration $u_1 -u_2$.
     
    At time $u_2$, the payoff from the options maturing at $u_2$ can be invested in the money market. Over the time period $[u_2, u_1]$, the portfolio then consists of two parts :
    \begin{enumerate}
        \item The initial portfolio of options with maturity $u_1$.
        \item The interest earned from the money market investment of the payoff from options maturing at $u_2$ as done for our simulations in sections \ref{DH-BS} and \ref{DH-MJD}.
    \end{enumerate}
    \ \item Instead of investing in the money market at time $u_2$, the agent can incorporate new options in the portfolio using the payoff from the options with short maturity $u_2$: At any time $t\in (u_2,u_1)$, if new options maturing at $u_3$ become available, where $u_2<t\leq u_3<u_1$, they can be included in the portfolio using equation (\ref{hedgeint4}), by simply replacing $u_2$ by $u_3$ in (\ref{hedgeint4}). \textbf{ The weights $w(\mathcal{K}_1)$ corresponding to the options with maturity $u_1$ remain unchanged.}
 \end{enumerate}
   \end{enumerate}
  \end{rem}

 \subsubsection{Application of Gaussian Quadrature and Gauss Laguerre to construct the hedging portfolio}\label{lag}
As mentioned earlier, trading takes place only over finite strike points and hence, the hedge portfolio thereby constructed has to be a finite sum instead of a continuum of short maturity calls. Therefore, to construct an equivalent hedging portfolio, each of the two integrals in (\ref{hedgeint4}) needs to be discretized to a finite sum, as done in \cite{carr2014static}. The corresponding expression for the first integral is then given by
 \begin{align*}
     \int_{K_{11}}^{K_{12}}w(\mathcal{K}_1)C(S,t,\mathcal{K}_1,u_1)d\mathcal{K}_1\approx\sum_{j=1}^{N}\mathcal{W}_{1j}(\mathcal{K}_{1j})C(S,t,\mathcal{K}_{1j},u_1),
 \end{align*}
 where the weights, $\mathcal{W}_{1j}$'s and the corresponding strikes, $\mathcal{K}_{1j}$'s are computed using the Gaussian Quadrature scheme as discussed in the Appendix.
 
The associated approximation error is
 \begin{align*}
&\int_{K_{11}}^{K_{12}}w(\mathcal{K}_1)C(S,t,\mathcal{K}_1,u_1)d\mathcal{K}_1-\sum_{j=1}^{N}\mathcal{W}_{1j}(\mathcal{K}_{1j})C(S,t,\mathcal{K}_{1j},u_1)\\
     &=\mathcal{O}\left(\frac{g^{2N}(\eta)}{(2N)!}\right),
 \end{align*}
with $g(x)=w(x)C(S,t,x,u_1)$ and for some $\eta \in (K_{11},K_{12})$.

 For approximating the first integral in (\ref{modifiedwght1}), one needs to perform Gaussian Quadrature twice, the inner one to compute the integral with respect to $\mathcal{K}_1$, over the interval $[0, K_{11}]$, which once obtained, is used to calculate the outer integral over $\mathcal{K}_2$, over the bounded interval $[K_{21},K_{22}]$.
 
 For the computation of the second integral in (\ref{modifiedwght1}), one needs to approximate the inner integral over $[K_{12},\infty]$ using a shifted Gauss-Laguerre integration and perform Gaussian Quadrature for the outer integral over $[K_{21},K_{22}]$. 

 Similar to the method of Gauss-Hermite quadrature, the Gauss-Laguerre quadrature method is used to approximate integrals of the form $\int_{0}^{\infty} e^{-x} f(x)dx $, for a sufficiently smooth function $f(x)$.  For a given target function $f (x)$, the Gauss-Laguerre quadrature rule generates a set of weights $w^l_{i}$ and nodes $x^l_{i}$, $i = 1,2,...N$, that are defined by
 \begin{align*}
     \int_{0}^{\infty} e^{-x} f(x)dx \approx \sum_{i=1}^{N} w^l_{i}f(x^l_{i}) +\frac{(N!)^2}{(2N)!}f^{(2N)}(\xi),
 \end{align*}
 for some $\xi \in (0,\infty)$.
 
 A shifted Laguerre method approximates an integral $\int_{a}^{\infty} e^{-x} f(x)dx $, where $a>-\infty$, for a sufficiently smooth function $f(x)$, by performing a change of variable to $x+a$ to the above integral to obtain the following approximation
 \begin{align}
 \label{gausslaguerre}
     \int_{a}^{\infty} e^{-x} f(x)dx \approx e^{-a}\sum_{i=1}^{N} w^l_{i}f(x^l_{i} +a) +\frac{(N!)^2}{(2N)!}f^{(2N)}(\xi_a),
 \end{align}
 for some $\xi_a \in (a,\infty)$. 
 The reader can refer to the Appendix for a detailed outline of the Gauss-Laguerre method performed for our integral at hand and refer to \cite{davis2007methods} for a detailed description of the Gauss-Hermite, Gauss-Laguerre as well as Gaussian Quadrature methods.

Stated below are the corresponding formulae for the weights (\ref{GQweights}) and (\ref{modifiedwght1}) for the $BS$ and $MJD$ models, which shall be used for all our numerical experiments in Section \ref{sec5}.
\subsection{Black-Scholes model} 
 Consider the $BS$ model where, under the risk-neutral framework, the stock price follows a Geometric Brownian Motion ($GBM$) given by
 \begin{align}
     dS_t =(r-\delta)S_tdt+\sigma S_t dW_t,
 \end{align}
 where $W_t\sim N(0,t)$ denotes the standard Wiener process.
 
 Equation (\ref{GQweights}) for obtaining the weights associated to the options with short maturity $u_1$ under the $BS$ model translates to
 \begin{align}
    w(x) =e^{-\delta(T-u)}\frac{n(d_1)}{x\sigma\sqrt{T-u}},
 \end{align}
 with
 \begin{align*}
     d_1=\frac{ln(\frac{x}{K})+(r - \delta +\frac{\sigma^2}{2})(T-u)}{\sigma\sqrt{T-u}},
 \end{align*}
 and
 \begin{align*}
     \label{call}
     C(S,t,x,u)=S e^{-\delta(u-t)}N(\hat{d}_1)-xe^{-r(u-t)}N(\hat{d}_2),
 \end{align*}
 with
 \begin{align*}
     &\hat{d}_1=\frac{ln(\frac{S}{x})+(r - \delta +\frac{\sigma^2}{2})(u-t)}{\sigma\sqrt{u-t}},\\
     &\hat{d}_2= \hat{d}_1 -\sigma\sqrt{u-t},
 \end{align*}
where $N(.)$ denotes the cdf of a standard normal random variable.

Under the $BS$ model, the modified weight $\tilde{w}_2(\mathcal{K}_2)$, given by equation (\ref{modifiedwght1}) and associated with options with short maturity $u_2$, would then be obtained by substituting
 \begin{align}
   w_2(\mathcal{K}_2,\mathcal{K}_1) = e^{-\delta(u_1 - u_2)}\frac{n(\hat{\hat{d}}_1)}{\mathcal{K}_2\sigma\sqrt{u_1 -u_2}},
 \end{align}
 with
 \begin{align*}
     \hat{\hat{d}}_1=\frac{ln(\frac{\mathcal{K}_2}{\mathcal{K}_1})+(r -\delta +\frac{\sigma^2}{2})(u_1-u_2)}{\sigma\sqrt{u_1 -u_2}}.
 \end{align*}
\subsection{Merton Jump Diffusion model}\label{sec4}
The Merton $(1976)$ Jump-diffusion ($MJD$) model is a Markovian model where the movements of the underlying asset price are modeled by
\begin{align}
    \label{MJD stock}
    \frac{dS_t}{S_t}=(r-\delta-\lambda g)dt+\sigma dW_t+dJ(\lambda),
\end{align}
with $dJ$ denoting a compound Poisson jump with intensity $\lambda$.

Conditional on a jump occurring, the log price follows a normal distribution with mean $\mu_j$ and variance $\sigma^2_j$, while the mean percentage price change is given by $g=(e^{\mu_j+\sigma^2_j/2}-1)$.

In the $MJD$ dynamics, the price of a European call option can be expressed as a weighted average of the $BS$ call pricing functions, with the weights being given by the Poisson distribution
\begin{align*}
\begin{split}
    C(S,t,K,T,\theta) &= e^{-r(T-t)}\sum_{n=0}^{\infty}Pr(n)[Se^{(r_n-\delta)(T-t)}N(d_{1n}(S,t,K,T))\\
&-KN(d_{1n}(S,t,K,T)-\sigma_n\sqrt{T-t})],
\end{split}
\end{align*}
where $Pr(n)$ refers to the probability mass function of a Poisson distribution and is given by
\begin{align*}
    Pr(n)=e^{-\lambda(T-t)}\frac{(\lambda(T-t))^n}{n!}.
\end{align*}
The function $d_{1n}(S,t,K,T)$ is defined as
\begin{align*}
    d_{1n}(S,t,K,T) =\frac{ln(S/K)+(r_n-\delta+\sigma^2_n/2)(T-t)}{\sigma_n\sqrt{T-t}},
\end{align*}
with
\begin{align*}
    r_n = r - \lambda g +n(\mu_j+\sigma^2_j/2)/(T-t),\\
    \sigma^2_n = \sigma^2 + n\sigma^2_j/(T-t).
\end{align*}

In the $MJD$ model, the delta and the strike weighting functions corresponding to the first short maturity $u_1$ are given by
\begin{align*}
    \Delta = e^{-2r(T-t)}\sum_{n=0}^{\infty}Pr(n)e^{r_n(T-t)}N(d_{1n}(S,t,K,T)),\\
    w(\mathcal{K})=e^{-r(T-u_1)}\sum_{n=0}^{\infty}Pr(n)e^{(r_n-\delta)(T-u_1)}\frac{n(d_{1n}(\mathcal{K},u_1,K,T)}{\mathcal{K}\sigma_n\sqrt{T-u_1}}.
\end{align*}

The strike points based on Gauss-Hermite quadrature $\{x_j,w_j\}_{j=1}^N$, as defined in \cite{carr2014static}, are
\begin{align*}
    \mathcal{K}_j = K e^{x_j\sqrt{2v(T-u_1)}+(\delta-r-v/2)(T-u_1)},
\end{align*}
where
\begin{align*}
    v =\sigma^2+\lambda((\mu_j)^2+\sigma^2_j),
\end{align*}
is the annualized variance of the asset return under the measure $\mathbb{Q}$. The corresponding portfolio weights are given by \cite{carr2014static}
\begin{align*}
    \mathcal{W}_j = \frac{w(\mathcal{K}_j)\mathcal{K}_j\sqrt{2v(T-u_1)}}{e^{-x^2_j}}w_j.
\end{align*}
\subsection{Application of Gaussian Quadrature to the MJD model}
The integrals in equation (\ref{hedgeint1}) can be computed for the $MJD$ model in an analogous manner as in $BS$ model, to obtain the modified weight (\ref{modifiedwght1}) using

\begin{align}
\label{modifiedwghtMJD}
    w_2(\mathcal{K}_2,\mathcal{K}_1)=e^{-r(u_1-u_2)}\sum_{n=0}^{\infty}\tilde{Pr}(n)e^{(\tilde{r}_n-\delta)(u_1-u_2)}\frac{n(\tilde{d}_{1n}(\mathcal{K}_2,u_2,\mathcal{K}_1,u_1))}{\mathcal{K}_2 \tilde{\sigma}_n \sqrt{u_1 -u_2}},
\end{align}
and
\begin{align}
\label{modifiedwghtMJD2}
    w(\mathcal{K}_1)=e^{-r(T-u_1)}\sum_{m=0}^{\infty}Pr(m)e^{(r_m-\delta)(T-u_1)}\frac{n(d_{1m}(\mathcal{K}_1,u_1,K,T))}{\mathcal{K}_1 \sigma_m \sqrt{T -u_1}},
\end{align}
with
\begin{align*}
    &\tilde{d}_{1n}(\mathcal{K}_2,u_2,\mathcal{K}_1,u_1)=\frac{ln(\mathcal{K}_2/\mathcal{K}_1)+(\tilde{r}_n-\delta+\tilde{\sigma}^2_n/2)(u_1-u_2)}{\tilde{\sigma}_n\sqrt{u_1-u_2}},\\
     &\tilde{Pr}(n)=e^{-\lambda(u_1-u_2)}\frac{(\lambda(u_1-u_2))^n}{n!},\\
     &\tilde{r}_n=r-\lambda g + n(\mu_j+\sigma^2_j/2)/(u_1-u_2),\\
     &\tilde{\sigma}^2_n=\sigma^2+n\sigma^2_j/(u_1-u_2),
\end{align*}
and
\begin{align*}
    &d_{1m}(\mathcal{K}_1,u_1,K,T)=\frac{ln(\mathcal{K}_1/K)+(r_m-\delta+\sigma^2_m/2)(T-u_1)}{\sigma_m\sqrt{T-u_1}},\\
    &Pr(m)=e^{-\lambda(T-u_1)}\frac{(\lambda(T-u_1))^m}{m!},\\
    &r_m=r-\lambda g + m(\mu_j+\sigma^2_j/2)/(T-u_1),\\
    &\sigma^2_m=\sigma^2+m\sigma^2_j/(T-u_1).
\end{align*}

Here, $\mathcal{K}_1$ and $\mathcal{K}_2$ correspond to the strike points obtained by application of the Gaussian Quadrature over the intervals $[K_{11}, K_{12}]$ and $[K_{21}, K_{22}]$ respectively.
 \section{Numerical results}\label{sec5}
In this section, we apply the Gaussian Quadrature method, discussed in detail in Section \ref{sec4}, for hedging a European call option and use calls with both one as well as two short maturities to construct the hedge. The key assumption is that the options corresponding to the short maturities $u_1$ and $u_2$  are available in the ranges $[K_{11}, K_{12}]$ and $[K_{21}, K_{22}]$ respectively.

Throughout the rest of the paper, we shall use the notations $GQ_1$ and $GQ_2$ to denote the Gaussian Quadrature hedges obtained using options with one and two short maturities respectively. The first part of this section is dedicated to a detailed analysis of the performance of the Gaussian Quadrature methods, $GQ_1$ and $GQ_2,$ along with the Carr-Wu method \cite{carr2014static}, at initial time $t_0=0$,  for the $BS$ and $MJD$ models. The experiments have been designed to depict the efficiency of our method when compared to the Carr-Wu method \cite{carr2014static} and thereby, highlight their practical significance.

 The only restriction that we impose while applying the Carr-Wu method \cite{carr2014static} for the purpose of our numerical experiments throughout this paper is that the strike points in the expression (\ref{carrwu2}) are restricted to be in the interval $[K_{11}, K_{12}]$, as done for our Gaussian Quadrature ($GQ_1$) method. We apply the Carr-Wu method in two ways to construct the hedge:
 \begin{enumerate}
     \item $CW_a$ denotes  the application of the method with the number of quadrature points, $N_a$ being chosen such that the corresponding strike points $K_1,..., K_{N_a}$, all lie in the interval $[K_{11}, K_{12}]$ 
     \item $CW_b$ denotes the application of the method with the number of quadrature points, $N_b$, being chosen to be the same as for $GQ_1$ and the strike points falling outside the interval $[K_{11}, K_{12}]$ are dropped.
 \end{enumerate}
 
For the second part of the numerical results, we present the performance of these methods at an intermediate time, under the $BS$ and $MJD$ models, using simulated stock paths. We report the following statistics: the $95$th percentile, $5$th percentile, root mean squared error ($RMSE$), mean, mean absolute error ($MAE$), minimum (Min), maximum (Max), skewness and kurtosis, when applied to $GQ_1, GQ_2, CW_a, CW_b$ and Delta Hedging ($DH$). Following \cite{carr2014static}, we have kept Delta Hedging as a benchmark for these numerical experiments and reported their corresponding statistics. In related literature, \cite{wu2016simple} also use Delta Hedging as a benchmark for their numerical experiments.

 For simplicity of notations, we assume a zero dividend rate $\delta =0$ in all our experiments for the $BS$ model. Delta Hedging is then performed using the following method:
 
 If $V_0(S_0)$ denotes the initial value of the hedge, then by the self-financing condition we have
 \begin{align*}
     V_0(S_0)=C(S_0,0,K,T).
 \end{align*}
 We then divide the time interval $[0,T]$ into finite number of equi-spaced time-points $0=t_0<t_1<...<t_n=T$, such that $\Delta t=t_{i+1}-t_i$, $i=0,..,n-1$.
 
 Then, by the Delta Hedging argument, the value of the hedge portfolio  at each time step $t_i$, $i>0$, is  given by
\begin{align*}
    V_i= \Delta_{i-1} S_i +(V_{i-1} - \Delta_{i-1} S_{i-1})e^{r\Delta t},
\end{align*}

 where $\Delta_i$ denotes the Greek delta of the call option at time $t_i$.

 \subsection{Black-Scholes Model:}
 \subsubsection{Effect of number of quadrature points}
 In the first experiment, we list the results obtained by hedging using the Carr-Wu method and the Gaussian Quadrature method, involving both one and two short maturities, as we keep varying the number of quadrature points for both methods.
 
For the first experiment, we do not include the options of shorter maturity $u_2$ since the errors for $GQ_1$, as seen in Table \ref{comparison1} are already low, so an introduction of a second short maturity is not necessary and would not affect the results.

 Table \ref{comparison1} reports the expected discounted loss ($EDL$) of the hedge at initial time $0$ when the hedge is constructed. The formula for the expected discounted loss is
 \begin{align}
 \begin{split}
     \label{EDL}
     EDL &= \text{value of target option at time}~ 0 \\
     &-~\text{value of the hedge portfolio at time 0}.
     \end{split}
 \end{align}
  The reason behind the terminology of $EDL$ is that it represents the portion of the risk that cannot be hedged at the initial time $0$ by the constructed hedging portfolio.
  
 The parameters used are: $ S_0=100,T=1, u_1=40/252\approx 0.1587, K=100,K_{11}=0,K_{12}=130,\sigma=0.27,\mu=0.1,r=0.06,\delta=0$. The value of the target call option is $13.5926277.$ 
 
 Thus, $u_1 =0.1587$ denotes a fraction of the target maturity, $T=1$. This would correspond to $58$ days if we consider the target maturity to be a year, constituting $365$ days. On the other hand, if we consider a year to constitute $252$ days, this would correspond to $40$ days. Further, the number in the brackets for $CW_b$ indicates the number of quadrature points falling in the range $[K_{11}, K_{12}]$.

  \begin{table}
\centering
 \begin{tabular}{|c|c|c|c|c|}
    \hline
    $N_a$ & $CW_a$ & $N_q$ & $CW_b$ & $GQ_1$   \\
    \hline  2 & 0.9464 & 50 & -0.00065(28) & -0.00067\\
            2 & 0.9464 & 25 & 3.2$e^{-5}$(15) &  -0.00067\\
            2 & 0.9464 & 15 & 0.00167(9) & -0.00067 \\
            2 & 0.9464 & 10 & -0.01357(6)& -0.00625 \\
           2 & 0.9464 & 8 & -0.01556(5) & -0.05559 \\
            2 & 0.9464 & 6 & -0.00568(4) & -0.28426 \\
         \hline
    \end{tabular}
    \vspace{0.2cm}
     \caption{Absolute-errors for $CW_a$, $CW_b$ and $GQ_1$ as the number of quadrature points are varied.} \label{comparison1}
\end{table}
In Table \ref{comparison1}, $N_a$ denotes the number of quadrature points used for applying $CW_a$ and $N_q$ denotes the number of quadrature points used for $CW_b$ and $GQ_1$ methods. For the given choice of parameter values, $N_a$ is restricted to $2$ since for higher values, some strike points lie outside the interval $[K_{11},K_{12}]$.

\begin{figure}[htbp]
 \centering

\begin{subfigure}[b]{.4\textwidth}
 \centering
  \includegraphics[width=.5\linewidth]{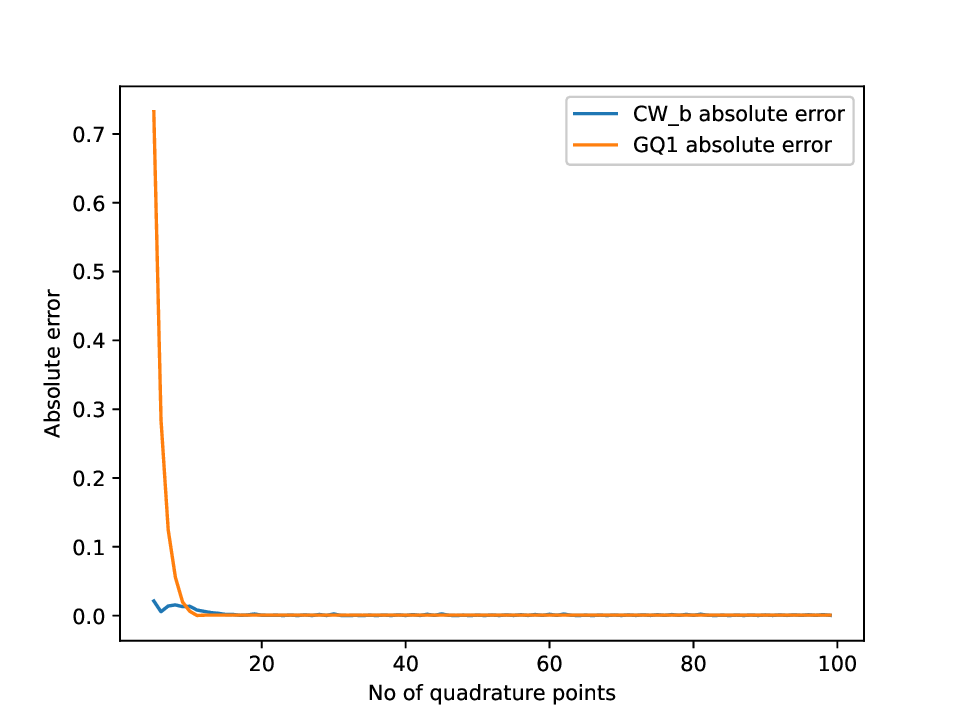}  
  \caption{ $[K_{11},K_{12}]=[0,130]$ }
  \label{fig:sub-first}
\end{subfigure}
\begin{subfigure}[b]{.4\textwidth}
 \centering
  \includegraphics[width=.5\linewidth]{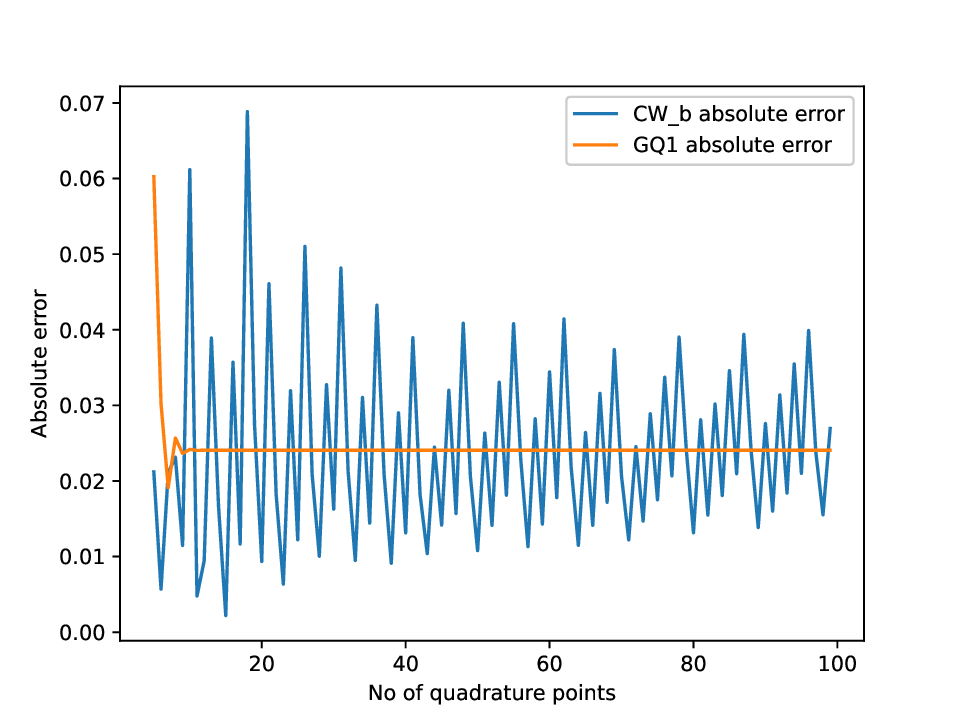}  
  \caption{ $[K_{11},K_{12}]=[40,130]$ }
  \label{fig:sub-second}
\end{subfigure}
\begin{subfigure}[b]{.4\textwidth}
 \centering
  \includegraphics[width=.5\linewidth]{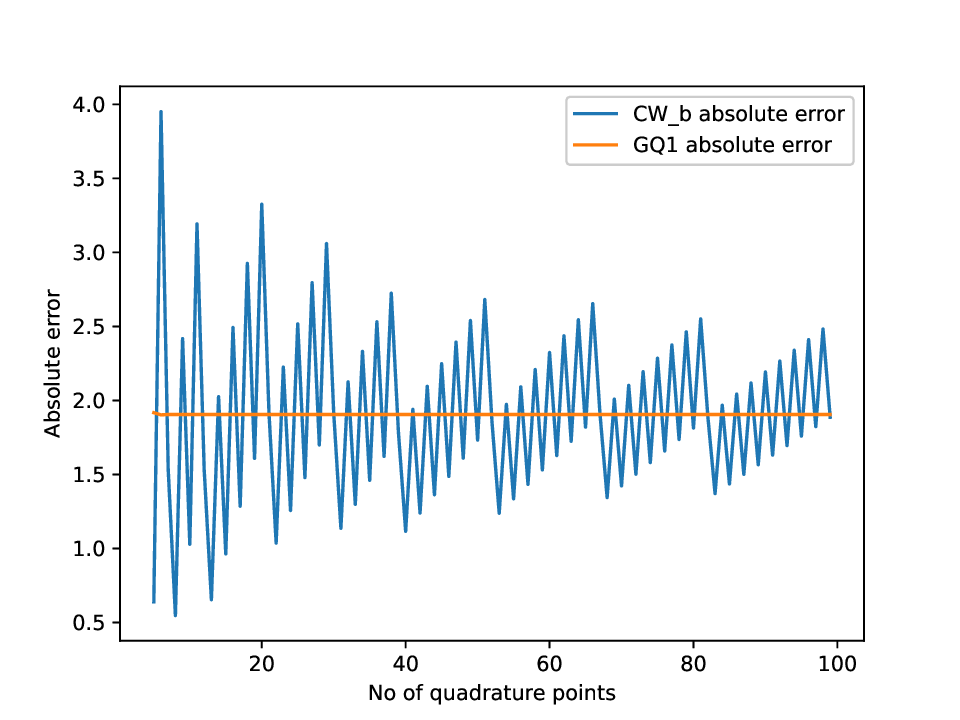}  
  \caption{  $[K_{11},K_{12}]= [60,130]$ }
  \label{fig:sub-third}
\end{subfigure}
\begin{subfigure}[b]{.4\textwidth}
 \centering
  \includegraphics[width=.5\linewidth]{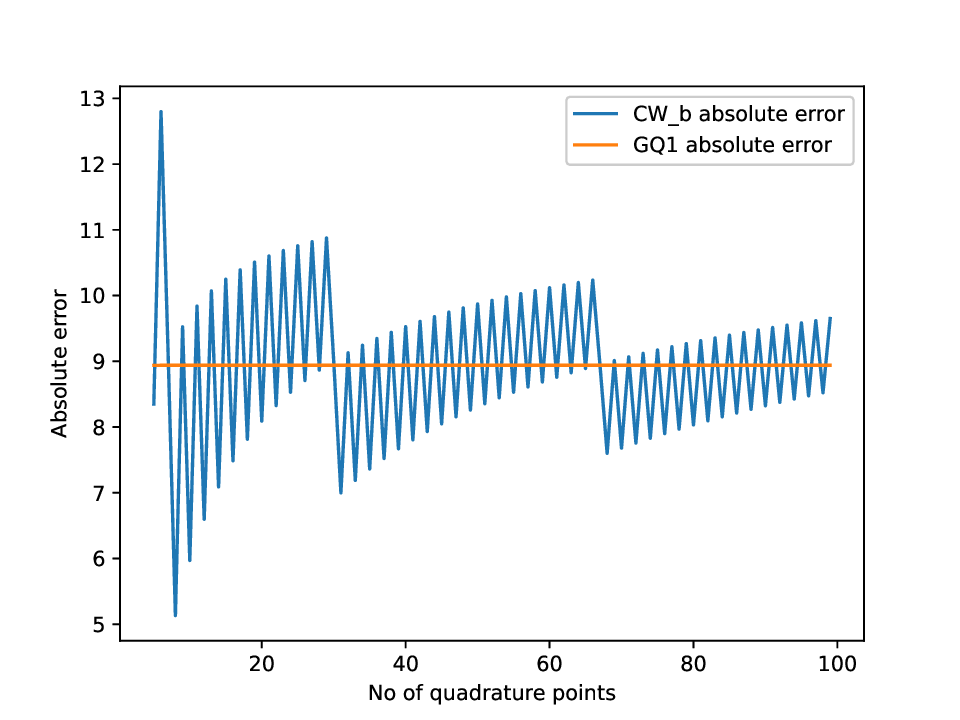}  
  \caption{ $[K_{11},K_{12}]=[80,130]$ }
  \label{fig:sub-fourth}
\end{subfigure}
\caption{Error plots for $CW_b$ and $GQ_1$ methods for increasing number of quadrature points, with $u_1= 0.1587$ and different strikes ranges, $[K_{11},K_{12}]$.}
\label{CW GQ_1 plots for varying quad}
\end{figure}

From the results listed in Table \ref{comparison1} and Figure \ref{CW GQ_1 plots for varying quad}, one can observe that the performance of $GQ_1$ improves as we keep increasing the number of quadrature points, up to a certain value of $N_q$, after which the performance becomes stable. Contrary to this, the performance of $CW_b$ fluctuates, sometimes to a large extent, depending on the strike points that fall in the range $[K_{11},K_{12}]$ and their associated weights.
 
This highlights the advantage of our Gaussian Quadrature hedging approach in obtaining a stable static hedge as we keep increasing the number of options used in constructing the hedge portfolio. When a sufficiently large number of strikes are available and the range of strikes is restricted, the Gaussian Quadrature method is more stable. On the other hand, $CW_b$'s performance would fluctuate in such a scenario.

To ensure simplicity of notations, for all future experiments, we use the same number of quadrature points ($N_q $) for both the short maturities $u_1$ and $u_2$. For calculating the modified weight (\ref{modifiedwght1}), we use $5$ and $20$ quadrature points for the application of the
Gaussian Quadrature and Gauss Laguerre methods respectively, which have been explained in detail in subsection \ref{lag}.

 \subsubsection{Effect of the range of strike intervals}
 
In this subsection we examine the effect of the restriction of the range of strike points, on the performance of the hedge, keeping the number of quadrature points to be fixed.

In an ideal scenario, when an investor witnesses high trade volumes in the market, where a large range of actively traded strikes are available, they can easily use either the Carr-Wu method or the Gaussian Quadrature method to construct their hedge portfolio and thereby, hedge the risk that they incur from short-selling the target call option. 

The problem arises when the strike range for actively traded options for a given maturity is then quite restricted. We suppose that the range of available strikes, $[K_{11},K_{12}]$ and $[K_{21},K_{22}]$ corresponding to the two short-maturities $u_1$ and $u_2$ respectively, is restricted. Further, our portfolio constitutes only $4$ options for $GQ_1$ and $CW_b$, and $4$ additional options with short maturity $u_2$, for the $GQ_2$ method.
  
Table \ref{tab:comparison2} lists the $EDL$ of the $CW_a, CW_b$, $GQ_1$ and $GQ_2$ methods. The strike points are restricted to the mentioned intervals. The strike points for $CW_a$ and $CW_b$ have been restricted over the interval $[K_{11}, K_{12}]$ and the number of quadrature points used for $CW_a$  and the actual number of strike points for $CW_b$ that fall in the strike interval $[K_{11}, K_{12}]$ have been mentioned in the brackets.

The inclusion of the second short maturity, assuming that the strikes for the second short maturity are in mentioned strike intervals ends up improving the hedging performance of the Gaussian Quadrature method as denoted by the percentage decrease in loss ($PDL$). The $PDL$ is calculated by the following formula
\begin{align}
    \label{PDL}
    PDL =~\frac{|\text{$EDL$ using $GQ_1$}|-~|\text{$EDL$ using $GQ_2$}|}{|\text{$EDL$ using $GQ_1$}|}\times 100\%.
\end{align}
The parameters used for the following experiment are : $ S_0=100,T=1,u_2=0.0833,u_1=0.1587,K=100,\sigma=0.27,\mu=0.1,r=0.06,\delta=0$. The value of the target call option is $13.5926277.$

 From Table \ref{tab:comparison2} one can notice that in certain cases holding the $CW_b$ or $CW_a$ hedge would provide better risk-exposure than $GQ_1$. It should be noted that one can further optimize the risk exposure using $GQ_2$ by including options with shorter maturities, $u_3, u_4,..,u_n$(say), with $u_n<...u_4<u_3<u_2<u_1$. 
 
Further, in the case of the $CW_a$ and $CW_b$ methods, the results would be highly dependent on the number of quadrature points used, as explained in the previous experiment. The Gaussian Quadrature, on the other hand, would provide stable results even in restricted strike intervals, after a certain number of quadrature points.
 
Table \ref{tab:comparison2} also highlights an important fact that a slight increase in the range of actively traded strikes corresponding to the second short maturity $u_2$ can have a substantial positive impact on the performance of the hedge. This performance can be improved by the addition of further short maturities $u_2>u_3>...>u_n>0$ by application of Corollary \ref{corollary}. 
\begin{table}[]
    \centering
    \begin{tabular}{|c|c|c|c|c|c|c|}
          \hline
     $[K_{11},K_{12}]$ & $[K_{21},K_{22}]$ & $CW_a$ & $CW_b$ & $GQ_1$ &$GQ_2$ &$PDL$  \\
    \hline  $[80,120]$ & $[80,120]$ & -3.8(1) &  -13.0(1) & -8.9 & -8.3 & 6.7\%\\
    $[80,120]$ & $[75,120]$ & -3.8(1) & -13.0(1) & -8.9 & -7.2 & 19.5 \%\\
    $[80,120]$ & $[55,120]$ & -3.8(1) & -13.0(1) & -8.9 & 1.6 & 82.2\%\\
    $[60,105]$ & $[60,105]$ & -3.8(1) &  -2.7(1) & -2.1 & -1.7 & 20.0\%\\
    $[75,110]$ & $[75,110]$ & -3.8(1) & -2.7(1) & -7.1 & -6.5 & 9.4\%\\
    $[55,110]$ & $[75,110]$ & -3.8(1) & -2.7(1) & -1.0 &  -0.9 & 6.7\%\\
    $[55,110]$ & $[65,105]$ & -3.8(1) & -2.7(1) & -1.0 & -0.9 & 4.7\%\\
     \hline
    \end{tabular}
    \vspace{0.2cm}
    \caption{$EDL$ comparison of $CW_b,GQ_1$ and $GQ_2$}
    \label{tab:comparison2}
\end{table}

 \subsubsection{Effect of the spacing between the target and the short maturities}
Let us consider the problem faced by the writer of a call option that matures in one year $(T=1)$ and is written at-the-money, as assumed in our previous example. The writer intends to hold this short position for an optimal time $u_1<T$, after which the option position will be closed. During this time, the writer can hedge their market risk using various exchange-traded assets such as the underlying stock, futures, and/or options on the same stock. In the case that the writer decides to hedge their position using options on the same stock, it is of utmost interest to compute the effect of the short maturities, $0<u_2<u_1<T$, on the performance of the hedge and accordingly minimize their risk exposure. 

Assuming enough trade volume in the market, we use $15$ quadrature points for computing the hedge portfolios for both $CW_b$ and the $GQ_1$ methods and $30$ quadrature points for the $GQ_2$ method. Further, we restrict the strike interval $[K_{11}, K_{12}]$ to a more realistic range to indicate the fact that actively traded short maturity options have strikes close to the target option's strike. The parameters are: $S_0=100,T=1,K=100,K_{11}=80,K_{12}=120, K_{21}=60,K_{22}=120,\sigma=0.27,\mu=0.1,r=0.06,\delta=0$. The value of the target call option is $13.5926277.$ 

Table \ref{comparisonmaturity1} reports the $EDL$ of the $CW_b$, $GQ_1$, and $GQ_2$ methods as we vary the short maturity $u_1$, while keeping the second short maturity fixed at $u_2 =0.0079$. 
\begin{table}
\centering
 \begin{tabular}{|c|c|c|c|c|c|c|}
    \hline
     $u_1$  & $N_a$ & $CW_a$ &$CW_b$ & $GQ_1$ & $GQ_2$ & $PDL$ \\
    \hline
    0.0833 & 1 & -4.2 & -10.6(2) & -9.6 & -2.0 & 78.6\%\\
    0.1587 & 1 & -3.8 & -10.2(2) & -8.9 & -2.5 & 72.3\% \\
    0.3175 & 1 & -2.8 & -9.6(2) & -7.5 & -2.4 & 67.3\%\\
    0.6349 & 1 & -1.3 & -3.6(3) & -3.8 & -1.4 & 63.6\%\\
         \hline
    \end{tabular}
    \vspace{0.2cm}
     \caption{$EDL$ for the $CW_a$, $CW_b$, $GQ_1$ and $GQ_2$ as the short maturity $u_1$ is varied, with strikes $[K_{11},K_{12}]=[80,120]$ and $[K_{21},K_{22}]=[60,120]$}.\label{comparisonmaturity1}
\end{table}
It can be inferred from Table \ref{comparisonmaturity1} that for an investor with a very restricted range of actively traded strikes at their disposal, the $GQ_2$ method would serve as a better method for minimizing their risk exposure.

It should also be noted from the last two rows of Table \ref{comparisonmaturity1} that even though $CW_a$ gives a comparable performance to $GQ_2$ in the case when $u_1$ is closer to the target maturity $T=1$, with only one strike point being used for $CW_a$, the results would vary considerably if the actual strike in the mentioned range $[80,120]$ is quite far away from the strike point given by $CW_a$. While for $GQ_2$ we have $15$ distinct choices of strike points in each of the intervals $[80,120]$ and $[60,120]$, so the actual strike points would be close to $GQ_2$ strike points.

One can further increase the quadrature points in $GQ_2$ to ensure that the actual strike points are very close to quadrature points (without impacting the results, owing to the stability of the $GQ_2$ method with increasing quadrature points, after a certain number of quadrature points) as shown in the first experiment, which is not the case for $CW_a$ or $CW_b$.

If, on the other hand, the range of actively traded strikes corresponding to the first short maturity $u_1$ is wide as given by the parameters: $S_0=100, T=1, K=100, K_{11}=60, K_{12}=120, K_{21}=60, K_{22}=120,u_2=0.0079,\sigma=0.27,\mu=0.1,r=0.06,\delta=0$, then, choosing the same number of quadrature points for $CW_b$, $GQ_1$ and $GQ_2$, as done in Table \ref{comparisonmaturity1}, one would obtain the results listed in Table \ref{comparisonmaturity2}. On observing the results in both Tables \ref{comparisonmaturity1} and \ref{comparisonmaturity2}, it can be concluded that the performance of the $GQ_2$ hedge improves as we keep increasing the short maturity $u_1$, keeping everything else fixed.
\begin{table}
\centering
 \begin{tabular}{|c|c|c|c|c|c|c|}
    \hline
     $u_1$  & $N_a$ & $CW_a$ & $CW_b$ & $GQ_1$ & $GQ_2$ & $PDL$ \\
    \hline
    0.0833 & 2 & 1.27 & -0.96(4) & -2.36 & -2.36 & 0.1\%\\
    0.1587 & 2 & 0.94 & -0.96(4) & -1.91 & -1.63 & 14.6\% \\
    0.3175 & 2 & 0.52 & -0.92(4) & -1.11 & -0.85 & 23.3\%\\
    0.6349 & 2 & 0.11 & -0.31(5) & -0.27 & -0.06 & 76.6\%\\
         \hline
    \end{tabular}
    \vspace{0.2cm}
     \caption{$EDL$ for the $CW$ and $GQ_1$ as the maturity spacing is varied, with strikes $[K_{11},K_{12}]=[K_{21},K_{22}]=[60,120]$.} \label{comparisonmaturity2}
\end{table}
 \begin{figure}[htbp]
     \centering
     \includegraphics[scale=0.5]{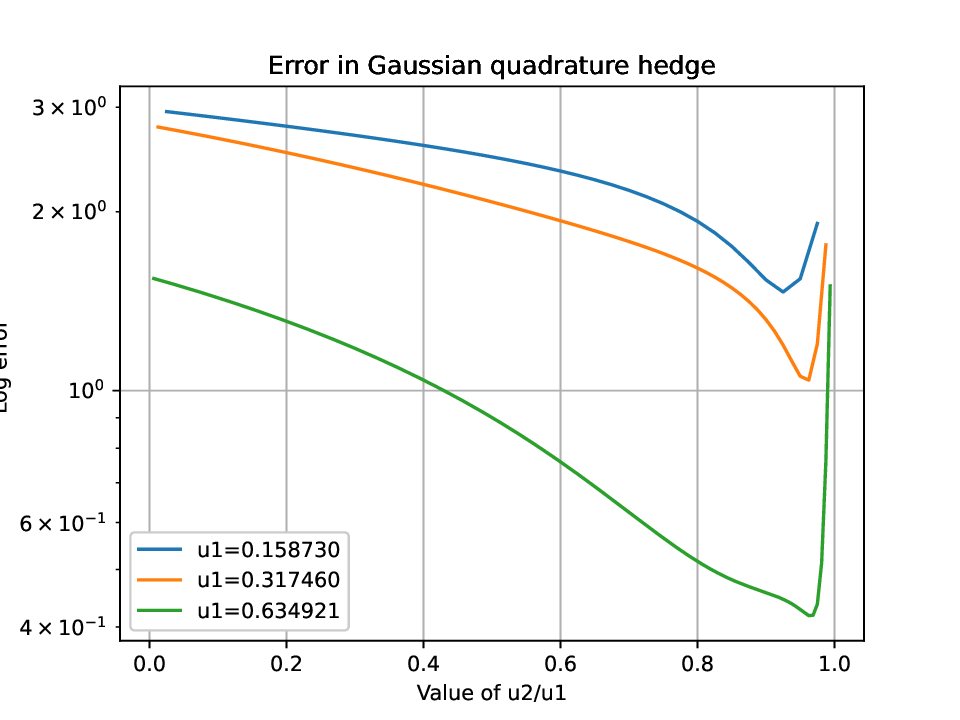}
     \caption{Log errors of the $GQ_2$ hedge as $u_2$ is varied, for $[K_{11},K_{12}]=[80,120]$ and $[K_{21},K_{22}]=[60,120]$}
     \label{fig:u2 varying error}
 \end{figure}

Figure \ref{fig:u2 varying error} displays the error in the $GQ_2$ hedge for three different choices of the first short maturity $u_1$, while increasing the short maturity $u_2$ to approach $u_1$ for each such choice. It can be concluded from Figure \ref{fig:u2 varying error} that the error in the $GQ_2$ hedge decreases as the second short maturity $u_2$ approaches $u_1$, with a sudden jump as $u_2$ gets extremely close to $u_1$. The jump arises due to the discontinuity in the call option pay-off at time $u_1$, owing to a factor of $u_2-u_1$ in the denominator for obtaining the modified weight given by equation (\ref{modifiedwght1}), associated with options with maturity $u_2$.

From a practical viewpoint, this implies that an investor should accumulate options of short maturities, with maturity dates close to each other to obtain significant improvements in the performance of his hedge, rather than just using one short maturity.

One should also note that, even if the short maturities are not close to each other, the resultant $GQ_2$ hedge with $N_1 + N_2$ options (say), would always have a better performance than that of the $GQ_1$ hedge constructed with only $N_1$ options. So, from an investor's perspective, it is always beneficial to include options of multiple short maturities in his hedge portfolio.
\subsubsection{Simulation based comparison with Delta Hedging}\label{DH-BS}
Following the series of experiments that have been done at the initial time $0$, the most natural thing to study would be to analyze the performance of the hedge until the expiry $u_1$ of the short maturity options. 

Since the $GQ_2$ hedge constitutes options with two short maturities, $0<u_2<u_1$, we incorporate the fact that at short maturity $u_2$, the payoff corresponding to the options with short maturity $u_2$ is invested in a risk-free bank account and the corresponding interest earned from this at every time $u_2<t\leq u_1$ is also a part of our hedging portfolio value at time $t$. 

The $EDL$  of the $CW_a, CW_b, GQ_1, GQ_2$ hedges at time $0$ are denoted by $B_0$. These are the approximation errors incurred due to the usage of a finite number of short-maturity options instead of the continuum of short-maturity options, given by the integrals in the corresponding hedge portfolios.

Depending on the sign, these errors are each invested in / borrowed from the money market at time $0$ and the interest incurred constitutes a part of the hedge portfolio error at each time $0<t_i\leq u_1$, as done in \cite{carr2014static}.

  We construct the hedging portfolio using two short maturities while simultaneously constructing the Delta Hedging portfolio. The Delta Hedging portfolio is rebalanced once at each of the equi-spaced time points over the interval $[0,u_1]$. We report the statistics at the time points, $u_2$ and $u_1$, respectively, corresponding to the maturities of the shorter-term options.
  
  For the Carr-Wu hedge portfolio, we only include the options with short maturity $u_1$ to emphasize the effect of the exclusion of shorter maturity $u_2$ on the performance of the hedge.

Table \ref{deltahedge} reports the $RMSE$ of the $CW_a, CW_b$, $GQ_1$ and $GQ_2$ methods at short maturity, $u_2$, with the strike points being restricted to the mentioned strike intervals. Table \ref{deltahedge2} in Appendix \ref{Appendix_DH_BS}, gives the corresponding errors at maturity $u_1$. To obtain the results, we simulate $1000$ stock paths, each at $N$ equispaced time-points $0<t_1<t_2..<t_{N} =u_1$,  with the spacing $t_{i}-t_{i-1}=h$ and report the  $RMSE$ for the three schemes.  

The parameters used for Tables \ref{deltahedge} and \ref{deltahedge2} are: $ S_0=100,T=1,u_2=0.0833,u_1=0.1587,h=0.004,N=40,K=100,K_{11}=80,K_{12}=120,K_{21}=60,K_{22}=120,\sigma=0.27,\mu=0.1,r=0.06,\delta=0$.

For the delta hedge, we rebalance the portfolio $40$ times, after equal intervals of $h=0.004$ each, where the target maturity is $T=1$. The modified weight (\ref{modifiedwght1}) associated with options with short maturity $u_2$ is estimated using $5$ and $20$ quadrature points, respectively. 

It can be concluded from Table \ref{deltahedge} that the performance of the $DH$ obtained by the frequent rebalancing of the portfolio is superior to the $CW_a, CW_b, GQ_1, GQ_2$. The performance of the $GQ_2$ method is considerably good but $DH$ still has an edge over this method, for the restricted range of strikes $[K_{11},K_{12}]=[80,120]$ and $[K_{21},K_{22}]=[60,120]$, corresponding to the options with short-maturity $u_1$ and $u_2$, respectively. Over the duration, $(u_2,u_1]$, when the options with short maturity $u_2$ have already expired, the hedge portfolio only consists of the options with maturity $u_1$ and the interest earned from the money market from the payoff of the shorter maturity options with maturity, $u_2$, as explained in Section \ref{multi}.

Later, in Table \ref{deltahedge_app_BS} of the Appendix \ref{Appendix_DH_BS}, we list the results for increased strike ranges in $[K_{11},K_{12}]$. It can be concluded from Table \ref{deltahedge_app_BS} that the performance of $GQ_1$ and $GQ_2$ greatly improve and can outperform the Delta Hedging performance when the strike ranges are wide enough.

Further, the Delta Hedging performance deteriorates rapidly when we consider jump-diffusion dynamics like the $MJD$ model, as shown in Section \ref{DH-MJD}.

\begin{table}
\centering
 \begin{tabular}{|c|c|c|c|c|c|}
    \hline
    Statistics  & $DH$ & $CW_a$&
     $CW_b$ &
      $GQ_1$ &
     $GQ_2$ \\
    \hline 
   No. of quad points & & 1  & 15(2)  & 15 & 15\\
    \hline
          $95$th percentile & 0.184 & 3.063 & 3.979 & 3.328 & 0.650 \\
          $5$th percentile & -0.209 & -1.698 & -4.792 & -3.996 & -0.773\\
          RMSE & 0.123 & 1.563 & 2.705 & 2.267 & 0.440\\
          Mean & 0.001 & 0.033 & -0.073 & -0.061 & -0.011\\
          MAE & 0.098 & 1.302 & 2.155 & 1.804 & 0.351\\
          Min & -0.498 & -1.712 & -9.242 & -8.033 & -1.585\\
          Max & 0.303 & 4.906 & 7.626 & 6.387 & 1.305\\
          Skewness & -0.516 &  0.798 & -0.304 & -0.325 & -0.291\\
          Kurtosis & 0.522 & -0.265 & -0.010 & 0.048 & 0.029\\
          \hline
    \end{tabular}
    \vspace{0.2cm}
     \caption{Comparison of hedging errors at short maturity $u_2$} \label{deltahedge}
\end{table}

\begin{figure}[htbp]
    \centering
    \includegraphics[scale=0.5]{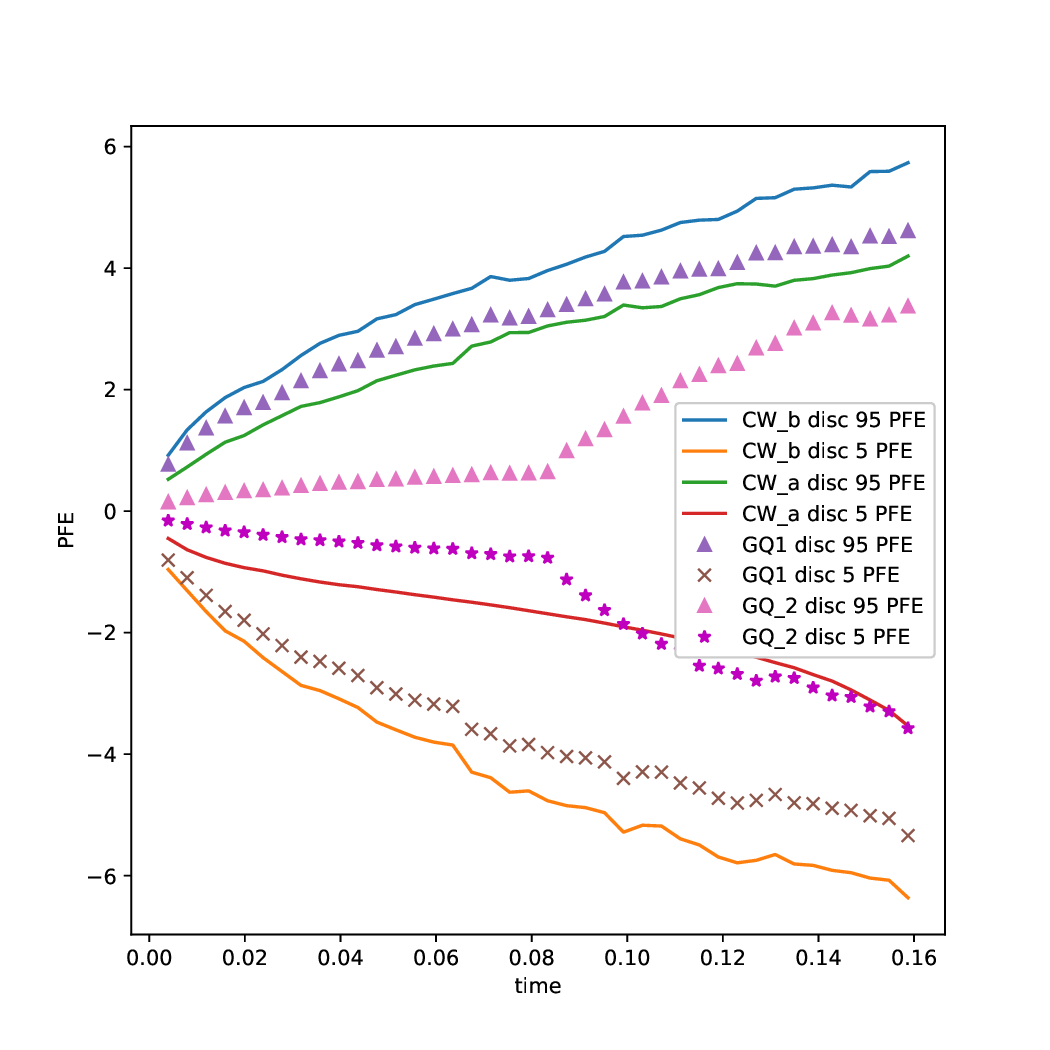}
    \caption{Plots of the discounted $95$th and $5$th percentiles of the various methods}
    \label{fig:PFA plots}
\end{figure}

Figure \ref{fig:PFA plots} displays the corresponding discounted $95$th and $5$th potential future exposures $(PFE)$ of the $CW_a, CW_b, GQ_1$ and $GQ_2$ methods for the parameters used in Table \ref{deltahedge}, till maturity $u_1$. \textbf{However, what is relevant here is the graph upto time $u_2$.} It can be observed from Figure \ref{fig:PFA plots} that the discounted $PFE$s of $GQ_2$ are significantly lower than the corresponding $PFE$s of $CW_a, CW_b$ and $GQ_1$ up to the second short maturity $u_2$, indicating better hedging of the investor's risk exposure up to time $u_2$ on including the options with short maturity $u_2$, which was the motivation behind including such options.

Figure \ref{fig:PFA plots} highlights an important factor. Over the time period $u_2<t\leq u_1$, if the investor invests the proceeds earned at the expiry of the options corresponding to short-maturity $u_2$ in a bank account, the hedge portfolio would still perform better overall compared to $CW_a, CW_b$, and $GQ_1$ portfolios. \textbf{While the discounted $5$-th percentile for the $CW_a$ method, given by the red line, is lower than the corresponding $5$-th percentile for the $GQ_2$ hedge, it is highly sensitive to the available strike points in the strike range $[K_{11}, K_{12}]$, as explained earlier.} 

By using our algorithm, the investor can also incorporate newly available options available at any time $t\in(u_2,u_1)$, with maturity $u_3\in(t, u_1]$, along with their already existing portfolio of options with short-maturity $u_1$, as explained in the Remark following Corollary \ref{corollary}. This would give a significant reduction in the hedging error.

\subsection{Merton Jump Diffusion model}
For the $MJD$ model, we shall repeat a similar sequence of experiments to the one done for the $BS$ model and report the corresponding results.

\subsubsection{Effect of the number of quadrature points}
 
 Table \ref{MJD quad} presents the results obtained at initial time $t_0 =0$ when the number of quadrature points is varied for $CW$ and $GQ_1$ while restricting the strike points of $CW$ to be in the range $[K_{11}, K_{12}]$. Since for $N_c>3$, some of the strike points obtained using $CW$ lie outside $[K_{11},K_{12}]$, we exclude such strike points.
 
 The parameters used  are: $ S_0=100, T=1, u_1=0.1587, K=100, K_{11}=0, K_{12}=150, \sigma=0.14, \mu=0.1, r=0.06, \delta=0.02, \sigma_j =0.13, \mu_j =-0.1,\lambda=2$. The value of the target call option is $11.9882525.$
\begin{table}
\centering
 \begin{tabular}{|c|c|c|c|c|}
    \hline
    $N_c$ & $CW_a$ & $N_q$ & $CW_b$ & $GQ_1$   \\
    \hline  3 & 1.47 & 5 & -0.80(4) &  6.27 \\
             3 & -2.24 & 10 & -0.04(7)  & -0.34 \\
           3 & -2.24 & 15 & 0.04(10) &  0.01\\
            3 & -2.24 & 25 &  0.01(16) & 1.67$e^{-5}$\\
            3 & -2.24 & 50 & 1.19$e^{-4}$(29)  & -8.98$e^{-6}$\\
            3 & -2.24 & 100 & -6.82$e^{-6}$(56) & -8.98$e^{-6}$\\
\hline
    \end{tabular}
     \caption{$EDL$ for the $CW_a$, $CW_b$ and $GQ_1$ as the number of quadrature points are varied.} 
     \label{MJD quad}
\end{table}

From Table \ref{MJD quad} one can observe similar results as for the $BS$ model, where the Gaussian Quadrature method's performance is stable with respect to increasing quadrature points (after a certain number of points).
\subsubsection{Effect of strike range}
Table \ref{MJD strike2} lists the absolute errors at time $0$ for both the $CW_a, CW_b, GQ_1$, and $GQ_2$ methods, as the strike ranges are varied while keeping the number of quadrature points to be fixed. The actual number of strike points for $CW_b$ which fall in the strike interval $[K_{11},K_{12}]$ has been mentioned in brackets. For $CW_a$ we restrict ourselves to include only the strike points which fall in the range $[K_{11},K_{12}]$.

 The parameters used for Table \ref{MJD strike2} are: $ S_0=100, T=1, u_1 = 0.1587, u_2 = 0.0833, K=100, \sigma=0.14, \mu=0.1, r=0.06, \delta=0.02, \sigma_j =0.13, \mu_j =-0.1,\lambda=2$.The value of the target call option is $11.9882525.$
\begin{table}
\centering
 \begin{tabular}{|c|c|c|c|c|c|c|c|c|}
    \hline
    $[K_{11},K_{12}]$ & $[K_{21},K_{22}]$ & $N_a$ & $CW_a$ & $N_q$ & $CW_b$ & $GQ_1$ & $GQ_2$  & $PDL$\\
    \hline  $[80,120]$ & $[80,120]$ & 1 & -2.54 & 20 & -6.33(2) & -6.80 &  -6.52 & 4.10\%\\
    $[80,120]$ & $[75,120]$ & 1 & -2.54 & 20 & -6.33(2) & -6.80 &  -4.86 & 28.5\%\\
    $[80,120]$ & $[60,120]$ & 1 & -2.54 & 20 & -6.33(2) & -6.80 & -1.21 & 82.21\%\\
    $[75,110]$ & $[75,110]$ & 1 & -2.54 & 20 & -6.33(2) & -4.64 & -4.37 & 14.49\%\\
   $[60,105]$ & $[60,105]$ & 1 & -2.54 & 20 & -0.20(4) & -0.61 & -0.52 & 14.49\%\\
   $[55,110]$ & $[75,110]$ & 1 & -2.54 & 20 &  -0.20(4) & -0.20 & -0.19 & 7.44\%\\
    $[55,110]$ & $[65,105]$ & 1 & -2.54 & 20 &  -0.20(4) & -0.20 & -0.19 & 6.09\%\\
  
    \hline
    \end{tabular}
     \caption{Absolute-errors for the $CW$, $GQ_1$ and $GQ_2$ as the strike ranges are varied.}\label{MJD strike2} 
\end{table}

On observing Table \ref{MJD strike2} one can draw similar conclusions as for the $BS$ model that if the strike range corresponding to the first short maturity $u_1$ is wide enough, with enough actively traded options, then one can choose either $CW_a, CW_b$ or $GQ_1$ to construct his hedge. 

The addition of the options with the second short maturity, $u_2$, always leads to a reduction in the hedging error, with the most significant decrease being when the strike range, $[K_{21},K_{22}]$ corresponding to the short maturity $u_2$ is wider than $[K_{11},K_{12}]$ for $u_1$.

\subsubsection{Effect of the spacing between the target and the short maturities}
Table \ref{MJD strike2} lists the absolute errors at time $0$ for both the $CW_a, CW_b$, and $GQ_1$ methods, as the short maturity $u_1$ are varied while keeping everything else fixed.

The actual number of strike points for $CW_b$, which fall in the strike interval $[K_{11}, K_{12}]$, have been mentioned in brackets. For $CW_a$ we restrict ourselves to include only the strike points which fall in the range $[K_{11}, K_{12}]$.

 The parameters used for Table \ref{MJD u1}  are : $ S_0=100, T=1, u_1 = 0.1587,  K=100, \sigma=0.14, \mu=0.1, r=0.06, \delta=0.02, \sigma_j =0.13, \mu_j =-0.1,\lambda=2, [K_{11},K_{12}]=[80,120], N_q =20$. The value of the target call option is $11.9882525.$
\begin{table}
\centering
 \begin{tabular}{|c|c|c|c|c|}
    \hline
    $u_1$ & $N_a$ & $CW_a$ & $CW_b$ & $GQ_1$ \\
    \hline  0.0833 & 1 & -3.18 &  -6.63(2) & -7.47\\
    0.1587 & 1 & -2.54 & -6.33(2) & -6.80\\
    0.3175 & 1 & -1.29 & -5.73(3) & -5.22\\
    0.6349 & 2 & 0.14 & -0.89(4) & -1.65\\
    \hline
    \end{tabular}
     \caption{Absolute-errors for the $CW$, $GQ_1$ and $GQ_2$ as the strike ranges are varied.}\label{MJD u1} 
\end{table}
\begin{figure}[htbp]
    \centering
    \includegraphics[scale=0.5]{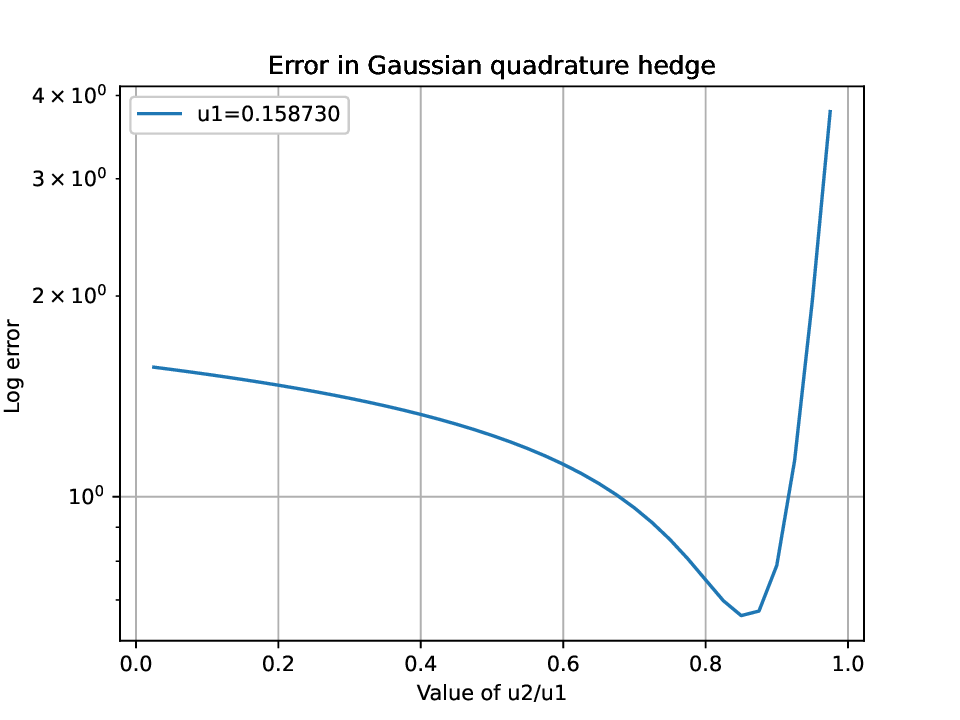}
    \caption{Error in $GQ_2$ hedge as $u_2$ is varied}
    \label{fig:MJD u2}
\end{figure}

Figure \ref{fig:MJD u2} plots the error in $GQ_2$ hedge as the second short-maturity $u_2$ approaches the first short maturity $u_1$, while keeping the other parameters fixed at : $ S_0=100, T=1, u_1 = 0.1587,  K=100, \sigma=0.14, \mu=0.1, r=0.06, \delta=0.02, \sigma_j =0.13, \mu_j =-0.1, [K_{11},K_{12}]=[80,120],[K_{21},K_{22}]=[60,120], N_q =20$. 

From Table \ref{MJD u1} and Figure \ref{fig:MJD u2}, we arrive at similar conclusions that the errors in the $GQ_1$ hedge are a monotonically decreasing function in short maturity $u_1$. In the case of $GQ_2$, the errors decrease until a certain time point close to the short maturity $u_1$, attain a minimum, and rapidly increase beyond that owing to the discontinuity, as in the case of the Black-Scholes model. 

The value of $u_2$ at which the minimum is attained for a given choice of parameters can be easily obtained by applying a simple bisection method. 
\subsubsection{Effect of distribution of jumps}
In this section, we would like to analyze the effect of changes in values of $\lambda,\mu_j$ and $\sigma_j$ on the performance of the $CW$ and $GQ_1$ hedges. 

We keep the annualized variance $v$ to be fixed at $0.27^2$ for each of the experiments. 

The reason for this study is to analyze the effect that the distribution of the jumps in the stock process would have on the hedging performance. 
\vspace{0.5cm}

\textit{Effect of change in $\lambda$:}  We study the effect of change in $\lambda$, while keeping $v,\mu_j$ and $\sigma_j$ fixed.  The values of $\lambda$ are chosen such that $\sigma =\sqrt{ v - \lambda(\mu^2_j + \sigma^2_j)}>0$.

The parameters used for Table \ref{MJD lambda} are: $ S_0=100, T=1, u_1 = 0.1587, K=100, K_{11} = 60 ,K_{12}=120, \mu=0.1, r=0.06, \delta=0.02, \sigma_j =0.13, \mu_j =-0.1$. 
\begin{table}
\centering
 \begin{tabular}{|c|c|c|c|c|c|}
    \hline
    $\lambda$ & $\sigma$ & $N_c$ & $CW_b$ & $N_q$ & $GQ_1$  \\
    \hline   0.02 & 0.2690 & 20 & 0.8117 & 20 & 1.5985\\
     0.1 & 0.2649 & 20 & 0.7796 & 20 & 1.5529\\
     0.5 & 0.2438 & 20 & 0.6239 & 20 & 1.3332\\
     1  & 0.2144 & 20 & 0.4435 & 20 & 1.0500\\
    \hline
    \end{tabular}
     \caption{Absolute-errors for the $CW_b$ and $GQ_1$ for increasing $\lambda$ , keeping the annualized variance fixed at $0.27^2$.}\label{MJD lambda} 
     
\end{table}

\begin{figure}[htbp]
\centering
\includegraphics[scale=0.5]{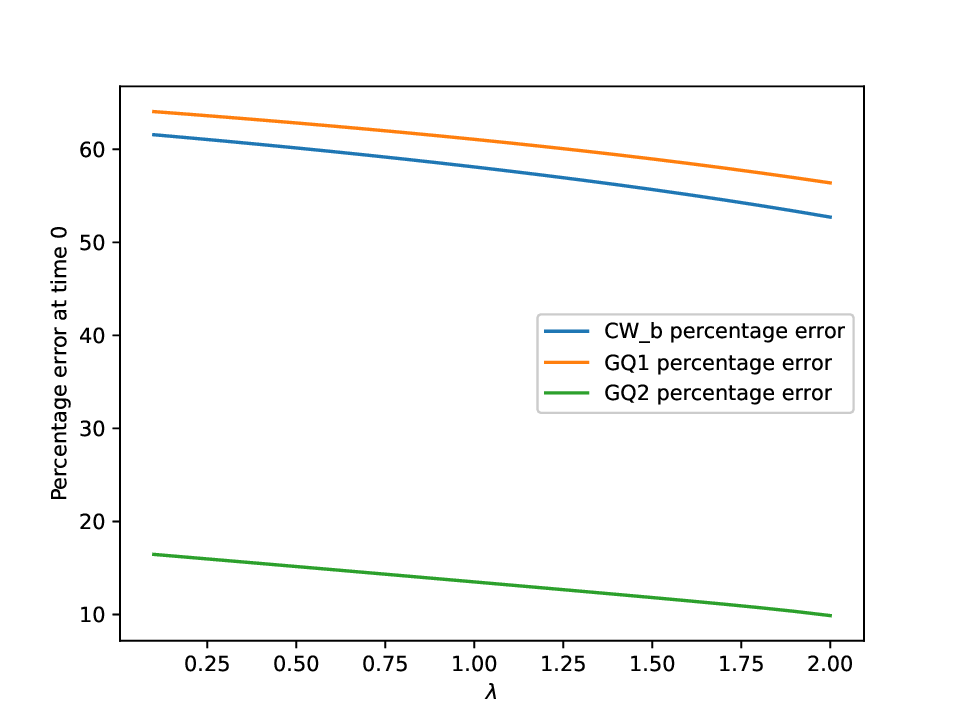}
\caption{ $CW_b$, $GQ_1$ and $GQ_2$ percentage error plots for varying $\lambda$, keeping the annualized variance fixed at $0.27^2$.}\label{fig: MJD lambda}
\end{figure}

The parameters used for  Figure \ref{fig: MJD lambda} are: $ S_0=100, T=1, u_1 = 0.1587,u_2=0.0833, K=100, K_{11}=80,K_{12}=120,K_{21} = 60 ,K_{22}=120, \mu=0.1, r=0.06, \delta=0.02, \sigma_j =0.13, \mu_j=-0.1$.

Note that the relation $\sigma=\sqrt{ v - \lambda(\mu^2_j + \sigma^2_j)}>0$ is a decreasing function of $\lambda$, when $v,\mu_j$ and $\sigma_j$ are fixed. It can be seen from Figure \ref{fig: MJD lambda} that the performance of $GQ_2$ is substantially better than that of $CW_b$ and $GQ_1$ in this scenario.
\vspace{0.5cm}

\textit{Effect of change in $\mu_j$: } We study the effect of change in $\mu_j$, keeping $\lambda,\sigma_j$ and $v$ fixed. The values of $\mu_j$ are chosen such that $\sigma >0$.

\begin{figure}[htbp]
\centering
\includegraphics[scale=0.5]{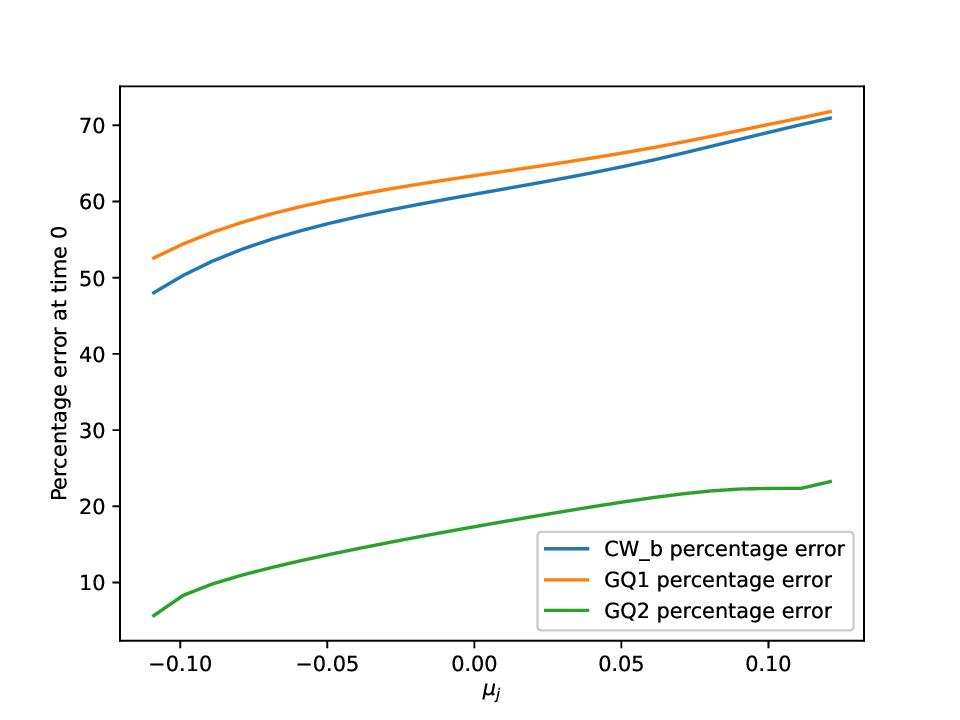}
\caption{ $CW$ and $GQ_1$ error plot for varying $\mu_j$ , keeping the annualized variance fixed at $0.27^2$.}\label{fig: MJD mu_j}
\end{figure}

The parameters used for  Figure \ref{fig: MJD mu_j} are: $ S_0=100, T=1, u_1 = 0.1587,u_2=0.0833, K=100, K_{11}=80,K_{12}=120,K_{21} = 60 ,K_{22}=120, \mu=0.1, r=0.06, \delta=0.02, \sigma_j =0.13, \lambda = 2$. 

The absolute error (at time $0$) of the $GQ_1$ increases with an increase in the average jump size $\mu_j$ (see Figure \ref{fig: MJD mu_j}), where $\lambda,\sigma_j$ and $v$ are the given constants and $\sigma=\sqrt{ v - \lambda(\mu^2_j + \sigma^2_j)}>0$.
\vspace{0.5cm}

\textit{Effect of change in $\sigma_j$: } We study the effect of change in $\sigma_j$ , while keeping $\lambda,\mu_j$ and $v$ fixed. The values of $\mu_j$ are chosen such that $\sigma >0$.

\begin{figure}[htbp]
\centering
\includegraphics[scale=0.5]{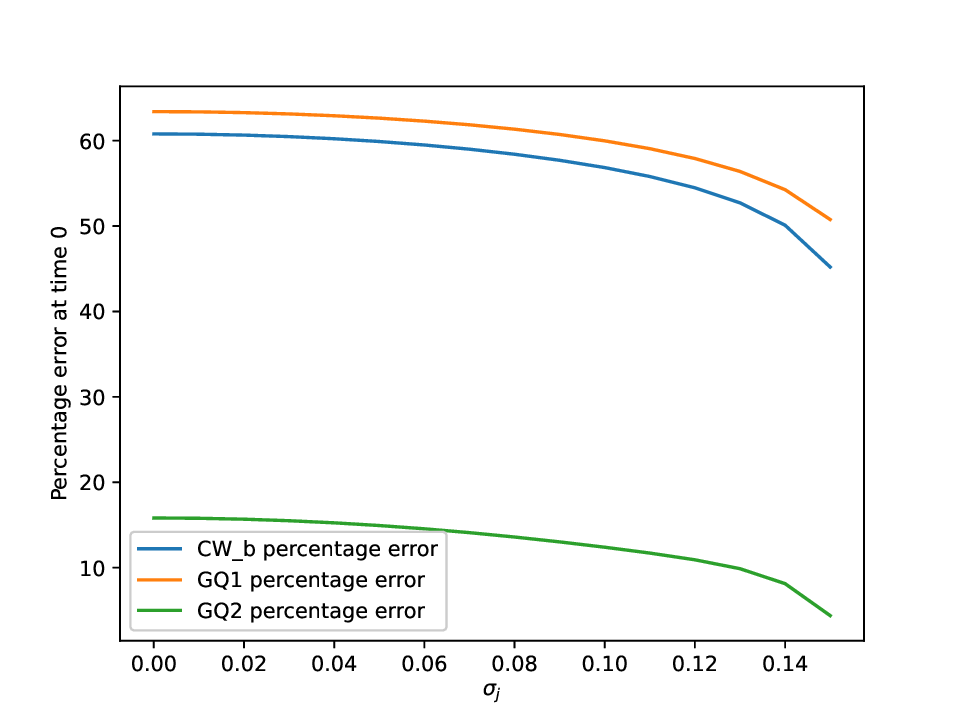}
\caption{ $CW_b$, $GQ_1$ and $GQ_2$ error plot for varying $\sigma_j$ , keeping the annualized variance fixed at $0.27^2$.}\label{fig: MJD sigma_j}
\end{figure}

The parameters used for Figure \ref{fig: MJD sigma_j} are: $ S_0=100, T=1, u_1 = 0.1587, u_2 = 0.0833, K=100, K_{11} = 80 ,K_{12}=120, K_{21} =60, K_{22}=120, \mu=0.1, r=0.06, \delta=0.02, \mu_j =-0.1, \lambda = 2$. 

As can be seen from Figure \ref{fig: MJD sigma_j}, $GQ_2$ performs better than that $CW_b$ and $GQ_1$, over the given restricted strike intervals.

\subsubsection{Simulation based comparison with Delta Hedging}\label{DH-MJD}
Table \ref{deltahedgeMJD1} reports the statistics of the $CW_a,CW_b,GQ_1$ and $GQ_2$ methods at time $u_2$, with the strike points restricted to the mentioned strike intervals. The parameters used for Table \ref{deltahedgeMJD1} are: $S_0=100,T=1,u_1=0.1587,u_2=0.0833,h=0.004,N=21,K=100,K_{11}=80,K_{12}=120,K_{21}=60,K_{22}=120,\mu=0.1,r=0.06,\delta=0.02,\sigma_j=0.13,\mu_j=-0.1,\lambda=2$.

The delta hedging portfolio is rebalanced $21$ times till the short maturity $u_2=0.0833$, at equal intervals of $h= 0.004$ each. The modified weights (\ref{modifiedwght1}) are estimated using $5$ and $20$ quadrature points, respectively. Since the options with short maturity $u_2$ are available till time $u_2=0.0833$, we have only included the results till short maturity $u_2$.

 \begin{table} 
\centering
 \begin{tabular}{|c|c|c|c|c|c|}
    \hline
    Statistics  & $DH$ & $CW_a$&
     $CW_b$ &
      $GQ_1$ &
     $GQ_2$ \\
    \hline 
   No. of quad points & & 1  & 20(2)  & 20 & 20\\
    \hline
          $95$th percentile & 0.246 & 2.432 &3.191 &3.721&0.467 \\
          $5$th percentile & -1.961 & -2.700 & -1.758 & -2.086 & -0.262\\
          RMSE & 1.219 & 1.560 & 1.472 & 1.647 & 0.219 \\
          Mean & -0.111 & -0.125 & 0.008 & 0.007 & 0.001\\
          MAE & 0.447 & 1.254 & 1.006 & 1.126 & 0.144 \\
          Min & -11.928 & -2.976 & -4.962 & -5.276 & -0.662\\
          Max & 0.277 & 6.053 & 6.353 & 6.820 & 1.212 \\
          Skewness & -5.508 & 0.282 & 1.522 & 1.407 & 1.873\\
          Kurtosis & 34.301 & -0.024 & 4.020 & 3.570 & 6.261 \\
          \hline
    \end{tabular}
    \vspace{0.2cm}
     \caption{Comparison of the hedging errors at short maturity $u_2$.} \label{deltahedgeMJD1}
\end{table}
It can be concluded from Table \ref{deltahedgeMJD1} that the performance of the $DH$ obtained by the frequent rebalancing breaks down in this case, with the maximum loss being almost twice that for $GQ_1$ and $18$ times for $GQ_2$. On the other hand, $GQ_2$ has far superior performance than any of the other methods $CW_a, CW_b$ and $GQ_1$, for the restricted range of strikes, $[K_{11}, K_{12}]=[80,120]$ and $[K_{21},K_{22}]=[60,120]$.

Table \ref{deltahedge_app_MJD} in Appendix \ref{Appendix_DH_MJD} lists the results for increased strike ranges in $[K_{11},K_{12}]$. The performance of $CW_b$ and $GQ_1$ greatly improve and are comparable to the performance of $GQ_2$.

\begin{figure}[htbp]
    \centering
    \includegraphics[scale=0.7]{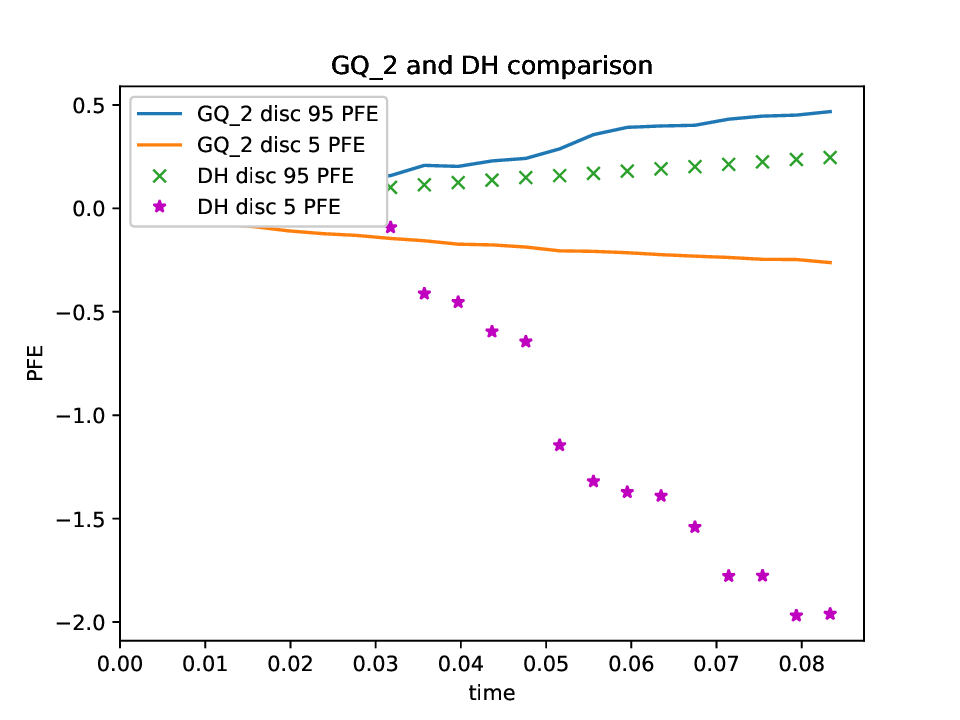}
    \caption{Plots of the discounted $95$th and $5$th percentiles of $GQ_2$ and $DH$.}
    \label{fig:PFA plots MJD}
\end{figure}

Figure \ref{fig:PFA plots MJD} displays the corresponding discounted $95$th and $5$th potential future exposures $(PFE)$ of the $GQ_2$ and $DH$ methods for the parameters used in Table \ref{deltahedgeMJD1}. It can be observed from Figure \ref{fig:PFA plots MJD} that the discounted $5$th PFE of $GQ_2$ is considerably lower than the corresponding $PFE$ of $DH$, indicating better hedging of the investor's risk exposure on including the options with the second short maturity, as desired.

\section{Conclusion}\label{sec6}
In this paper we have extended the theoretical spanning relation in \cite{carr2014static} to include options with multiple shorter maturities through Theorem \ref{theorem} and Corollary \ref{corollary}. An approximation of the exact spanning relation is then obtained by an application of the Gaussian Quadrature rule, as explained in detail in Section \ref{sec4} and Appendix. Numerical experiments are then performed in Section \ref{sec5} for the $BS$ and $MJD$ models lead to the following conclusions:
\begin{enumerate}
    \item The efficiency of the $GQ_1$ and $GQ_2$ methods can be increased as one keeps increasing the number of options held in the hedging portfolio up to a threshold, after which the performance stabilizes. The performance of $CW_b$ would fluctuate in such a scenario.
    \item  In case of restricted strike ranges, the inclusion of the second short maturity $u_2$, by application of the $GQ_2$ method improves the hedging performance when compared to both $GQ_1$ and $CW_a$ or $CW_b$. This improvement is substantial when the range of strikes available for the short-maturity $u_2$ is wider than that for the first short-maturity $u_1$.
    \item As observed for the Carr-Wu method, the closer the short-maturities are to the target option's maturity, $T$, the better the performance is for both the $GQ_1$ and $GQ_2$ methods. Further, the performance of the $GQ_2$ hedge improves as the spacing between the shorter maturities $u_1$ and $u_2$ keeps reducing.
    \item On the expiry of the options corresponding to the second short-maturity $u_2$, the investor has two choices at hand- $(i)$ They can invest the payoffs of these options in a bank account and continue with the initial portfolio corresponding to the options with short-maturity $u_1$. $(ii)$ They can choose to reinvest their payoffs to buy newly available options of other shorter maturities. The initial portfolio corresponding to the options with short-maturity $u_1$ stays intact in both $(i)$ and $(ii)$.
    
    In either case, the overall performance of the $GQ_2$ would be better than both the $CW_a$ and $CW_b$ methods, over restricted strike ranges. 
\end{enumerate}
While the results obtained in this paper illustrate the utility of our method from a hedging perspective, it is restricted to Markovian dynamics. Hence, as a natural extension of this work, extending this result for non-Markovian settings would serve as an important problem.

\section{Declarations of Interest:}
This work has been partially supported by the DST FIST program- $2021$ [TPN - $700661$]. The authors report no conflicts of interest.
\section{Acknowledgements}
The first author dedicates this work to the memory of their father.
\noindent
\bibliography{references.bib}
\newpage
\section{Appendix}\label{appendix}

\subsection{Approximation of an integral using Gaussian Quadrature rule}
There are various numerical schemes ranging from the Trapezoidal and Simpson's rule to more sophisticated ones over the recent past, for approximation of integrals over a bounded interval. While these numerical schemes have subtle differences among themselves, the general form of these approximation schemes is given as follows
\begin{align*}
    \int_{a}^{b}f(x)dx\approx A_0 f(x_0) +A_1 f(x_1) +...+A_n f(x_n),
\end{align*}
where
\begin{align*}
    &f(x)~~ \text{is the function whose integral needs to be approximated},\\
    &x_0,x_1,...x_n~~\text{are the nodes},\\
    &A_0,A_1,...,A_n~~\text{are the corresponding weights}.
\end{align*}
While in the Trapezoidal and Simpson's rules, the approach is to fix the nodes $x_i$'s, using which the weights $A_i$'s are found, the Gaussian Quadrature rule allows us to estimate both $x_i$'s and $A_i$'s, as dependent variables. The idea behind this approach is to choose $x_i$'s and $A_i$'s in a manner such that
\begin{align}
\label{gauss}
    \int_{a}^{b}f(x)dx\approx A_0 f(x_0) +A_1 f(x_1) +...+A_n f(x_n),~~\forall f\in \mathcal{P}_m,
\end{align}
where $\mathcal{P}_m$ denotes the vector space of polynomials of degree $\leq m$, where $m$, which denotes the degree of precision of the method, can be taken as large as possible.

The first observation that needs to be made in this regard is that for (\ref{gauss}) to hold, it is enough to show that the same holds for the basis functions: $ 1,x,x^2,...,x^m$, of the space $\mathcal{P}_m$.

This results in a set of $m+1$ equations which need to be solved for $2(N+1)$ unknowns, $A_i$'s and $x_i$'s, $i=0,1,2,...,N$, such that $m+1=2(N+1)$, which is simply the consistency condition.

In order to explain the idea better, let us first consider an example in the space $\mathcal{P}_3$. We wish to approximate the following integral
\begin{align}
\label{ex}
    \int_{-1}^{1} f(x)dx=A_0 f(x_0)+A_1 f(x_1),~~\forall f\in \mathcal{P}_3.
\end{align}
Hence, our task is now to check that (\ref{ex}) holds for $f(x):1,x,x^2,x^3$. An extremely useful formula in this regard is as follows
\begin{align*}
    \int_{-1}^{1}x^k dx=
    \begin{cases}
    \frac{2}{k+1},~~k~~\text{is even}\\
    0,~~k~~\text{is odd}.
    \end{cases}
\end{align*}
On substituting $f(x):1,x,x^2,x^3$ in (\ref{ex}) and utilising the above result we obtain the following system of equations
\begin{align*}
    &f(x)=1\Rightarrow 2=A_0+A_1,\\
    &f(x)=x\Rightarrow 0=A_0x_0+A_1x_1,
    &f(x)=x^2\Rightarrow \frac{2}{3}=A_0 x_{0}^2+A_1 x_{1}^2\\
    &f(x)=x^3\Rightarrow 0=A_0 x_{0}^3+A_0 x_{1}^3.
\end{align*}
This system can be easily solved to obtain the following values

\begin{align*}
    A_0=1,x_0=\frac{1}{\sqrt{3}};A_1=1,x_1=-\frac{1}{\sqrt{3}}.
    \end{align*}
If on the other hand, one wishes to approximate the following integral
\begin{align*}
     \int_{a}^{b} f(x)dx=\tilde{A}_0 f(t_0)+\tilde{A}_1 f(t_1),~~\forall f\in \mathcal{P}_3.
\end{align*}
then the desired nodes $t_i$'s and weights $\tilde{A}_i$'s in the interval $[a,b]$ can be obtained from the above obtained nodes, $x_i$'s and the corresponding weights $A_i$'s on $[-1,1]$,using the following linear transformations
\begin{align*}
    &t_i=\frac{1}{2}(b-a)x_i+\frac{1}{2}(a+b),\\
    &\tilde{A}_i=\frac{1}{2}(b-a)A_i.
\end{align*}
The most interesting fact about this approach is that the nodes lie in symmetric positions around the centre of the interval $[a,b]$ and correspondingly the weights assigned for each pair of symmetric points are the same, as can be seen in the example above.

\subsection{Delta Hedging results for $BS$ Model}\label{Appendix_DH_BS}
The parameters used for Table \ref{deltahedge2} are: $ S_0=100,T=1,u_2=0.0833,u_1=0.1587,h =0.004,N=40,K=100,K_{11}=80,K_{12}=120,K_{21}=60,K_{22}=120,\sigma=0.27,\mu=0.1,r=0.06,\delta=0$.
\begin{table}
\centering
 \begin{tabular}{|c|c|c|c|c|c|}
    \hline
    Statistics  & $DH$ & $CW_a$&
     $CW_b$ &
      $GQ_1$ &
     $GQ_2$ \\
    \hline 
   No. of quad points & & 1  & 15(2)  & 15 & 15\\
    \hline
          $95$th percentile & 0.294 & 4.237 & 5.788 & 4.658 & 3.405 \\
          $5$th percentile & -0.293 & -3.569 & -6.425 & -5.391 & -3.609\\
          RMSE & 0.175 & 2.422 & 3.716 & 3.102 & 2.137\\
          Mean & 0.007 & 0.037 & -0.010 & -0.012 & 0.037\\
          MAE & 0.142 & 2.044 & 2.943 & 2.467 & 1.695\\
          Min & -0.554 & -4.027 & -15.647 & -14.058 & -7.731\\
          Max & 0.427 & 6.159 & 9.235 & 7.921 & 6.409\\
          Skewness & -0.262 &  0.251 & -0.332 & -0.369 & -0.267\\
          Kurtosis & -0.168 & -0.894 &  0.055 & 0.245 &  0.248\\
          \hline
    \end{tabular}
    \vspace{0.2cm}
     \caption{Comparison of the hedging errors for $BS$ model at short maturity $u_1$.} \label{deltahedge2}
\end{table}

The parameters used for Table \ref{deltahedge_app_BS} are: $ S_0=100,T=1,u_2=0.0833,u_1=0.1587,h= 0.004,N=21,K=100,\sigma=0.27,\mu=0.1,r=0.06,\delta=0$.

\begin{table}
\centering
 \begin{tabular}{|c|c|c|c|c|c|}
 \hline
  \multicolumn{6}{|c|}{\textbf{Strikes: $[K_{11},K_{22}]=[60,120],[K_{21},K_{22}]=[60,120]$}} \\
    \hline
    Statistics  & $DH$ & $CW_a$&
     $CW_b$ &
      $GQ_1$ &
     $GQ_2$ \\
    \hline 
   No. of quad points & & 2  & 15(4)  & 15 & 15\\
    \hline
          $95$th percentile &0.184 & 0.888& 0.204&0.494&0.403\\
          $5$th percentile &-0.209&-1.920&-0.293&-0.617&-0.478\\
          RMSE & 0.123&0.903&0.151&0.349&0.273\\
    \hline
     \multicolumn{6}{|c|}{\textbf{Strikes:} $[K_{11},K_{22}]=[50,140],[K_{21},K_{22}]=[60,120]$} \\
    \hline
    Statistics  & $DH$ & $CW_a$&
     $CW_b$ &
      $GQ_1$ &
     $GQ_2$ \\
    \hline 
   No. of quad points & & 2  & 15(6)  & 15 & 15\\
    \hline
          $95$th percentile &0.184 & 0.888&0.023&0.079&0.079 \\
          $5$th percentile &-0.209&-1.920&-0.045&-0.094&-0.094\\
          RMSE & 0.123&0.903&0.021&0.053&0.053\\
          \hline
    \multicolumn{6}{|c|}{\textbf{Strikes:} $[K_{11},K_{22}]=[50,140],[K_{21},K_{22}]=[50,140]$} \\
    \hline
    Statistics  & $DH$ & $CW_a$&
     $CW_b$ &
      $GQ_1$ &
     $GQ_2$ \\
    \hline 
   No. of quad points & & 2  & 15(6)  & 15 & 15\\
    \hline
          $95$th percentile &0.184 & 0.888&0.023&0.079& 0.064\\
          $5$th percentile &-0.209&-1.920&-0.045&-0.094&-0.076\\
          RMSE & 0.123&0.903&0.021&0.053&0.043\\
          \hline
    \end{tabular}
    \vspace{0.2cm}
     \caption{Comparison of hedging errors for $BS$ model at short maturity $u_2$ for increased strike ranges.} \label{deltahedge_app_BS}
\end{table}
\subsection{Delta Hedging results for $MJD$ Model}\label{Appendix_DH_MJD}

The parameters used for Table \ref{deltahedge_app_MJD} are $S_0=100,T=1,u_1=0.1587,u_2=0.0833,h=0.004,N=21,K=100,\mu=0.1,r=0.06,\delta=0.02,\sigma_j=0.13,\mu_j=-0.1,\lambda=2$.
\begin{table}
\centering
 \begin{tabular}{|c|c|c|c|c|c|}
 \hline
  \multicolumn{6}{|c|}{\textbf{Strikes: $[K_{11},K_{22}]=[60,120],[K_{21},K_{22}]=[60,120]$}} \\
    \hline
    Statistics  & $DH$ & $CW_a$&
     $CW_b$ &
      $GQ_1$ &
     $GQ_2$ \\
    \hline 
   No. of quad points & & 1  & 20(4)  & 20 & 20\\
    \hline
          $95$th percentile &0.246&2.433&0.136&0.190&0.169\\
          $5$th percentile &-1.969&-2.699&-0.050&-0.107&-0.094\\
          RMSE & 1.190&1.560&0.062&0.087&0.078\\
    \hline
     \multicolumn{6}{|c|}{\textbf{Strikes:} $[K_{11},K_{22}]=[50,140],[K_{21},K_{22}]=[60,120]$} \\
    \hline
    Statistics  & $DH$ & $CW_a$&
     $CW_b$ &
      $GQ_1$ &
     $GQ_2$ \\
    \hline 
   No. of quad points & & 2  & 20(6)  & 20 & 20\\
    \hline
          $95$th percentile & 0.246&0.609&0.094&0.014&0.014\\
          $5$th percentile &-1.969&-1.376&-0.052&-0.008&-0.008\\
          RMSE & 1.190& 0.702&0.050&0.007&0.007\\
          \hline
           \multicolumn{6}{|c|}{\textbf{Strikes:} $[K_{11},K_{22}]=[50,140],[K_{21},K_{22}]=[50,140]$} \\
    \hline
    Statistics  & $DH$ & $CW_a$&
     $CW_b$ &
      $GQ_1$ &
     $GQ_2$ \\
    \hline 
   No. of quad points & & 2  & 20(6)  & 20 & 20\\
    \hline
          $95$th percentile & 0.246&0.609&0.094&0.014&0.013\\
          $5$th percentile &-1.969&-1.376&-0.052&-0.008&-0.007\\
          RMSE & 1.190& 0.702&0.050&0.007&0.006\\
          \hline
    \end{tabular}
    \vspace{0.2cm}
     \caption{Comparison of hedging errors for $MJD$ model at short maturity $u_2$ for increased strike ranges.} \label{deltahedge_app_MJD}
\end{table}
\end{document}